\documentclass[
aps,prd,
preprintnumbers,
amsmath,
amssymb,
nofootinbib
]{revtex4}
\usepackage{here}
\usepackage{footnote}
\usepackage{comment,braket}
\usepackage{epsf}
\usepackage{amsmath}
\usepackage{graphics}
\usepackage{amsfonts}
\usepackage{amssymb}
\usepackage{latexsym}
\usepackage{color}
\usepackage{natbib}
\usepackage{graphicx}
\usepackage{hyperref}
\usepackage[svgnames]{xcolor}
\definecolor{phthaloblue}{rgb}{0.0, 0.06, 0.54}
\hypersetup{
    colorlinks=true,
    linkcolor=black,
    citecolor=blue,
    filecolor=blue,
    urlcolor=black,
    }

\begin{document}
\title{Quantum Black Hole Seismology I: Echoes, Ergospheres, and Spectra}
\author{Naritaka Oshita$^{1}$}
\author{Daichi Tsuna$^{2,3}$}
\author{Niayesh Afshordi$^{1,4,5}$}
\affiliation{
  $^1$Perimeter Institute, 31 Caroline St., Waterloo, Ontario, N2L 2Y5, Canada
}

\affiliation{
  $^2$Research Center for the Early Universe (RESCEU), the University of Tokyo, Hongo, Tokyo 113-0033, Japan
}
\affiliation{
  $^3$Department of Physics, School of Science, the University of Tokyo, Hongo, Tokyo 113-0033, Japan
}

\affiliation{
  $^4$Department of Physics and Astronomy, University of Waterloo,
200 University Ave W, N2L 3G1, Waterloo, Canada
}

\affiliation{
  $^5$Waterloo Centre for Astrophysics, University of Waterloo, Waterloo, ON, N2L 3G1, Canada
}
\preprint{RESCEU-1/20}
\begin{abstract}

Searches for gravitational wave echoes in the aftermath of mergers and/or formation of astrophysical black holes have recently opened a novel and surprising window into the quantum nature of their horizons. Similar to astro- and helioseismology, study of the spectrum of quantum black holes provides a promising method to understand their inner structure, what we call {\it quantum black hole seismology}. We provide a detailed numerical and analytic description of this spectrum in terms of the properties of the Kerr spacetime and quantum black hole horizons, showing that it drastically differs from their classical counterparts. Our most significant findings are the following: (1) If the temperature of quantum black hole is $\lesssim 2 \times$ Hawking temperature, then it will not suffer from ergoregion instability (although the bound is looser at smaller spins). (2) We find how quantum black hole spectra pinpoint the microscopic properties of quantum structure. For example, the detailed spacing of spectral lines can distinguish whether quantum effects appear through compactness (i.e., exotic compact objects) or frequency (i.e., modified dispersion relation). (3) We find out that the overtone quasinormal modes may strongly enhance the amplitude of echo in the low-frequency region. (4) We show the invariance of the spectrum under the generalized Darboux transformation of linear perturbations, showing that it is a genuine covariant observable.
\end{abstract}

\maketitle

\section{Introduction}
Since the establishment of quantum mechanics and general relativity in the last century, constructing a theory of quantum gravity that integrates these two has been one of the biggest open questions in physics. Although numerous models of quantum gravity have been proposed, the major uncertainty is what occurs at the Planck scale. Therefore, the best way to test the models of quantum gravity is to find observable phenomena in which the Planckian physics can be probed. 

Recently, gravitational-wave (GW) echoes have been attracting much attention as a probe of the Planckian physics that may appear near the black hole (BH) horizons. Examples of such new physics include firewalls \cite{Almheiri:2012rt}, membranes \cite{Thorne:1986iy},  Gravastars \cite{Mazur:2001fv}, Fuzzballs \cite{,Mathur:2005zp},  and Exotic Compact Objects (ECOs) \cite{Schunck:2003kk,PhysRevLett.61.1446}, along with many other possibilities \cite{Holdom:2016nek,Barcelo:2014cla,Barcelo:2010vc,PrescodWeinstein:2009mp}. Two of the authors also recently pointed out that the modified dispersion relations at high frequencies may lead to the echo signals after the formation of BHs \cite{Oshita:2018fqu}. The common feature in all these models is that, in contrast to general relativity, which only has an ingoing wave solution at the BH horizon, the signal partially \textit{reflects} near the horizon, generating quasiperiodic signals with a period dependent on the BH's mass and spin \cite{Cardoso:2016rao,Cardoso:2016oxy,Abedi:2016hgu}. In general, the nonzero reflectivity of a BH could suffer from the ergoregion instability \cite{Nakano:2017fvh} that is caused by an infinite amplification due to the superradiance and nonzero reflectivity near the would-be horizon. This makes spinning BHs unstable. To avoid this, the reflectivity should be smaller than around $0.648$ for which the ergoregion instability is quenched up to $\bar{a} = 0.99999$ \cite{1974ApJ...193..443T,PhysRevD.99.064007}.

Although there are many quantum gravitational models which may lead to GW echoes, almost all of the theoretical echo analyses have been done under the assumption of \textit{constant reflectivity}, that is, the reflectivity near the horizon is assumed to be independent of the incoming frequency or the BH spin. Recently, two of the authors and Wang proposed the Boltzmann reflectivity model \cite{Oshita:2019sat,Wang:2019rcf} in which the energy reflectivity is given by $e^{-|\tilde{\omega}|/T_{\text{H}}}$, where $\tilde{\omega}$ is the frequency of incoming GWs in the horizon frame and $T_{\text{H}}$ is the Hawking temperature.
It is important to accurately predict the observational features of GW echoes from these models, not only to understand which model is correct, but also to maximize the possibility of finding these possible signals.

In this paper, we study the observational signatures of echoes for constant reflectivity, as well as generalized Boltzmann reflectivity models, where quantum horizon temperature may differ from $T_{\text{H}}$. We present in detail the modeling of GW echoes from spinning BHs, which can be used to obtain the outgoing echo spectrum from an arbitrary initial signal \cite{Mark:2017dnq,Wang:2019rcf}. While the term {\it black hole spectroscopy} is widely used (e.g., \cite{Dreyer:2003bv}), it often refers to quasinormal mode spectra of classical black holes that probe the classical geometry near the light ring. In contrast, we opt to call our study {quantum black hole seismology}, as  modes of quantum black holes live near their would-be horizons and probe their inner quantum structure (following similar applications of the term of ``seismology'' for stars and planets). 

The organization of this paper is as follows: In the next section, we introduce the constant and Boltzmann reflectivity models and summarize the constraints on their reflectivities from the ergoregion instability. In Sec. \ref{sec:template}, the details of the template for GW echoes in Kerr spacetime are presented\footnote{The GW echoes from a Kerr BH was investigated in \cite{Conklin:2019fcs,Wang:2019rcf} and the analytic approximation is discussed in \cite{Maggio:2019zyv}.} based on the Chandrasekhar-Detweiler (CD) \cite{Chandrasekhar:1976zz} and Sasaki-Nakamura (SN) equations \cite{10.1143/PTP.67.1788}. In Sec. \ref{sec:echo_spectrum}, we discuss two echo models based on the ECO  and the modified dispersion relation scenarios and investigate the difference between them in the spectrum of GW echoes. We also investigate the effect of overtone quasinormal modes (QNMs) and how the echo spectrum depends on the spin of BH and on the phase shift of GWs at the would-be horizon. In the latter part of this section, we show the invariance of the product of the reflectivity at the would-be horizon and the reflection coefficient of the angular momentum barrier under the generalized Darboux transformation, which guarantees the invariance of the peak frequencies in the echo spectrum under the generalized Darboux transformation.
The final section is devoted to conclusions.

In a forthcoming paper \cite{Oshita:seis2}, we investigate the consistency between the generalized Boltzmann reflectivity model and the tentative echo detection of GW170817 \cite{Abedi:2018npz} and discuss the detectability of echo signals from stellar collapse or failed supernovae, based on our methodology provided in this paper. Throughout the manuscript, we use the natural unit $\hbar=c=1$, and we denote the mass and angular momentum of a Kerr BH as $M$ and $a \ (\equiv \bar{a}GM)$, respectively, where $G\equiv 1/M_{\text{Pl}}^2$ is the gravitational constant and $M_{\rm Pl}$ is the Planck mass.

\section{Models and Constraints on the reflectivity}
\label{sec:model}
The amplitude and typical frequency of echoes depend on the reflectivity ${\cal R}$, the phase shift at the would-be horizon $\delta_{\text{wall}}$, and the reflection radius $x_0$. In this work, we investigate GW echoes with the following two models:
\begin{align}
{\mathcal R} =
\begin{cases}
R_c e^{i \delta_{\text{wall}}} & \text{constant reflectivity model,}\\
\exp\left({-  \displaystyle \frac{|\tilde{\omega}|}{2 T_{\text{QH}}}} +i \delta_{\text{wall}} \right) & \text{Boltzmann reflectivity model},
\end{cases}
\label{ref_con_bol}
\end{align}
where $R_c$ is a real constant, $\tilde{\omega} \equiv \omega -m  \Omega_H$, $\Omega_H \equiv \frac{\bar{a}}{2GM (1+\sqrt{1-\bar{a}^2})}$ is the horizon frequency, $m$ is the azimuthal number, $T_{\rm QH}$ is the Quantum Horizon temperature, and $\delta_{\text{wall}}$ is the phase shift at the would-be horizon. 

This general form for Boltzmann reflectivity was first introduced in \cite{Oshita:2018fqu}, as a result of the modified dispersion relation for gravitational waves near BH horizons:
\begin{equation}
\tilde{\Omega}^2 = \tilde{K}^2 +i \eta \tilde{\Omega} \tilde{K}^2 -C_d^2 \tilde{K}^4, 
\label{dispersion}
\end{equation}
where a reflectivity of the form (\ref{ref_con_bol}) with the following was shown:
\begin{equation}
T_{\rm QH} = \frac{\pi (1+4 C_d^2/\eta^2)}{\sqrt{2 + 4 C_d^2 /\eta^2}} T_{\rm H},
\label{quartic_temperature}
\end{equation}
where $C_{d}$ and $\eta$ are parameters that control the dispersion and dissipation effects, respectively, and $\tilde{K}$ and $\tilde{\Omega}$ are proper frequency and proper wave number of GWs, respectively, in the frame corotating with the horizon.
The analytic form in (\ref{quartic_temperature}) provides a good approximation to the solutions to the modified wave equation for $C^2_d \gg \eta$. In contrast, \cite{Oshita:2019sat} provides general arguments for $C_{d} = 0$ and $\gamma > 0$, which leads to the Boltzmann reflectivity of quantum horizons with $T_{\rm QH} = T_{\rm H}$. Of course the quantum gravity is still unknown physics and so we do not know the specific form of the Planckian correction to the dispersion relation. Therefore, in this manuscript, we treat the quantum horizon temperature as a parameter that, in general, differs from the Hawking temperature
\begin{equation}
T_{\rm QH} \neq T_{\rm H} \equiv \frac{1}{4 \pi G M} \left( \frac{\sqrt{1-\bar{a}^2}}{1+\sqrt{1-\bar{a}^2}} \right).
\end{equation}
In other words, we generalize the classical geometry near the would-be horizon to allow for a temperature that differs from the standard Hawking temperature for the quantum structure that reflect GWs. This makes sense as the original Hawking temperature relies on classical geometry near BH horizons \cite{Hawking:1974rv} that can be modified in a quantum theory of BHs. 

We also consider two types of the reflection radius $x_0$ with respect to the tortoise coordinate
\begin{align}
x_0 =
\begin{cases}
x_{\text{m}} &\equiv - \frac{1}{2 \pi T_H} \log{\left( \frac{M}{\gamma M_{\text{Pl}}} \right)},{\rm~for~ECO}\\
x_{\text{f}} &\equiv - \frac{1}{2 \pi T_H} \log{\left( \frac{M_{\text{Pl}}}{\gamma |\tilde{\omega}|} \right)},{\rm ~~for~modified~dispersion~relation}
\end{cases}
\label{x_m}
\end{align}
where $\gamma$ is a constant which determines the energy scale of exotic physics responsible for the reflection\footnote{Ref. \cite{Oshita:2018fqu} investigated the GW echoes induced by the modified dispersion relation (\ref{dispersion}) and found the relation between $\eta$ and $\gamma$, which is given by $\gamma \simeq (4C_d^2 +\eta^2)/\sqrt{2C_d^2+\eta^2}$.}. If reflection happens at the surface of an ECO, which stands at $\gamma \times$ proper Planck length outside the would-be event horizon, then $x_0 = x_{\text{m}}$, independent of the frequency of incoming GWs. On the other hand, in the modified dispersion relation model \cite{Oshita:2018fqu,Oshita:2019sat,Wang:2019rcf}, the reflection happens when the blue-shifted horizon-frame frequency of incoming GWs approaches Planck frequency $/\gamma$. However, in Sec. \ref{sec:x_0} we will show that there is only small difference between them except near $\omega= m \Omega_H$. Throughout the manuscript, we assume $\gamma = 1$, which is equivalent to assuming that the relevant energy scale at the would-be horizon is the Planckian energy. The fundamental frequency of GW echoes, $f_{\text{echo}}$, is given by
\begin{equation}
f_{\text{echo}} = \frac{1}{2 |x_0|},
\label{eq:fecho}
\end{equation}
which defines the quanta of the quansinomral mode frequencies of the quantum black holes \cite{Wang:2019rcf}. 

\begin{figure}[t]
    \includegraphics[width=0.55\textwidth]{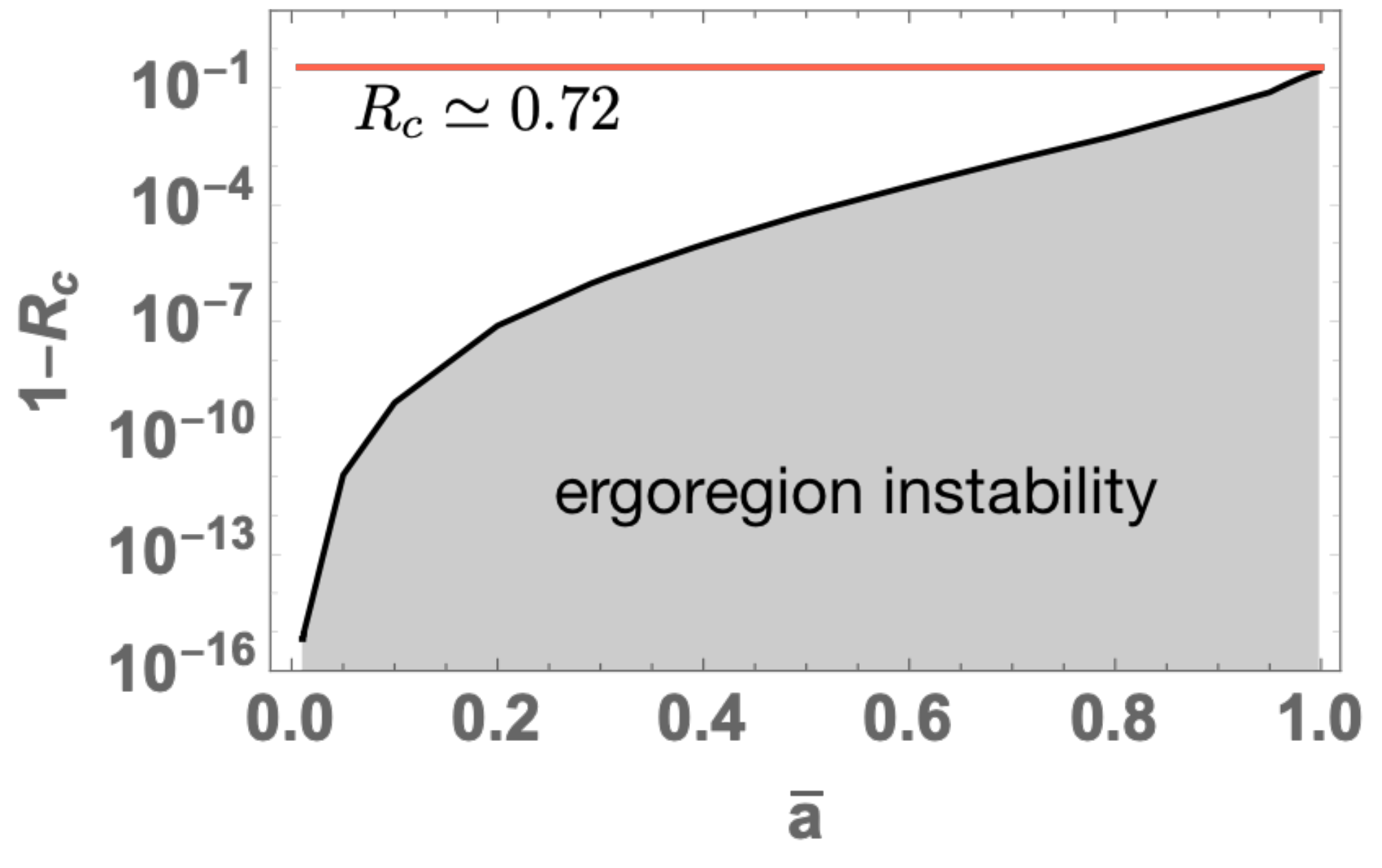}
\caption{Constraint on $R_c$ from the ergoregion instability up to the Thorne limit $0 \leq \bar{a} \leq 0.998$.
}
\label{C_super}
\end{figure}

\begin{figure}[h]
    \includegraphics[width=0.5\textwidth]{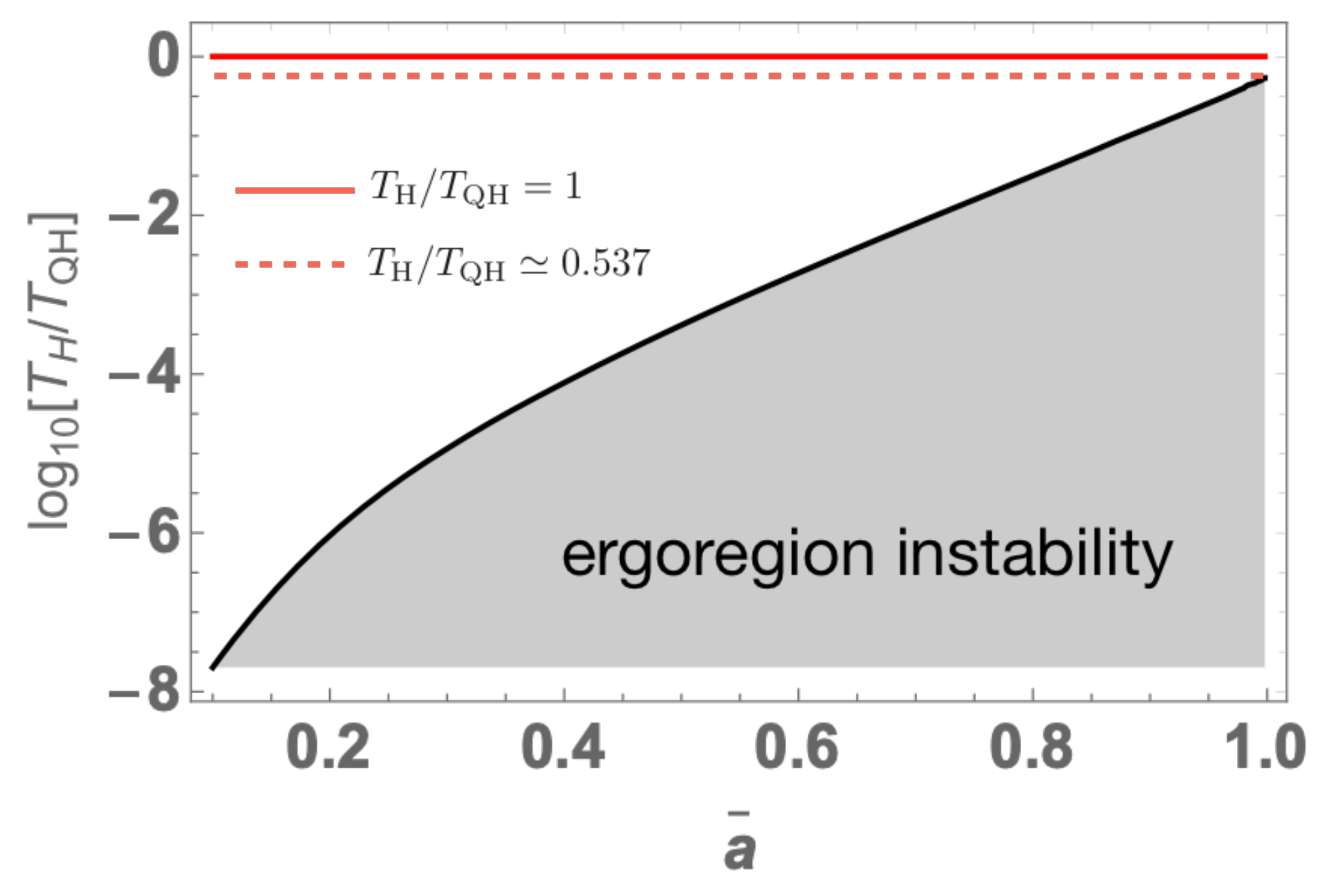}
\caption{Constraint on $T_{\text{H}} / T_{\text{QH}}$ from the ergoregion instability. One can read that $T_{\text{H}} / T_{\text{QH}} = 1$ has no instability up to the Thorne limit $\bar{a} \leq 0.998$ \cite{1974ApJ...191..507T}.
}
\label{B_super}
\end{figure}

Let us now summarize our findings for ergoregion instability for different reflectivity models: As we discussed in the Introduction, to avoid the ergoregion instability, an upper bound should exist for the constant reflectivity $R_c$. The numerical calculation of the superradiance predicts that the maximum amplification at $\bar{a}=0.998$ is around $0.91$, and so if $R_c$ has no spin dependence, it should be smaller than $ \frac{1}{\sqrt{1+0.91}} \simeq 0.72$ in order to avoid the instability in the range of $0 \leq \bar{a} \leq 0.998$ (see FIG. \ref{C_super}),\footnote{The amplification factor is $1.38$ for the near-extremal case $\bar{a}=0.99999$ \cite{1974ApJ...193..443T}, and so if such a BH exists in the Universe, the maximum reflectivity should be around $0.64$ to quench the ergoregion instability \cite{PhysRevD.99.064007}.} where the upper bound of $\bar{a} \leq 0.998$ is known as the Thorne limit \cite{1974ApJ...191..507T}. Allowing the spin dependence of $R_c$, the constraint on $R_c$ can be relaxed. However, as far as we know, there is no concrete theoretical model to predict a constant reflectivity and there is no specific prediction for the phase shift at the would-be horizon. In this manuscript, we therefore simply treat the phase shift $\delta_{\text{wall}}$ as a parameter.
We then turn to the more physical model provided by the Boltzmann reflectivity of Eq. (\ref{ref_con_bol})  \cite{Oshita:2018fqu,Oshita:2019sat,Wang:2019rcf}.
In FIG. \ref{B_super}, we show the constraint on the ratio $T_{\rm H} / T_{\rm QH}$ to avoid the ergoregion instability. We see that, in order to suppress the ergoregion instability with generalized Boltzmann reflectivity, the quantum horizon temperature should satisfy $T_{\rm QH} \lesssim 1.86 \times T_{\rm H}$.
Therefore, if only a dissipation term [the second term in Eq. (\ref{dispersion})] exists in the modified dispersion relation, the reflectivity is given by a Boltzmann factor with respect to the Hawking temperature $T_{\rm H} = T_{\rm QH}$ \cite{Oshita:2019sat}, which does not lead to the instability.
Using Eq. (\ref{quartic_temperature}), the bound for the quantum horizon temperature can be translated into $C_d^2 \lesssim 0.5 \gamma^2$ for the modified dispersion relation (\ref{dispersion}).

\section{Template for GW-echoes in Kerr spacetime}
In this section, we provide the details of the computation of a template for GW echoes in Kerr spacetime by using the (real wave) CD  and (complex wave) SN equations. GW echoes from a Kerr BH were investigated in \cite{Conklin:2019fcs,Wang:2019rcf,Conklin:2019smy} and the analytic approximation is discussed in \cite{Maggio:2019zyv}.
\label{sec:template}

\subsection{Wave equations in the Kerr spacetime and boundary conditions}
The spacetime around a Kerr BH with its mass $M$ and angular momentum of $a M$ is described by the following metric
\begin{equation}
ds^2=-\left(1-{r_sr\over \Sigma} \right) dt^2+{\Sigma\over\Delta}dr^2+\Sigma d\theta^2
+\left(r^2+a^2+{r_sra^2\over\Sigma}\sin^2{\theta} \right)\sin^2{\theta}d\phi^2-{{2r_sra\sin^2{\theta}\over\Sigma}}dt d\phi, 
\label{Kerr}
\end{equation}
where
$\Sigma \equiv r^2+a^2\cos^2{\theta}$ and $\Delta \equiv r^2-2GMr+a^2=r^2-r_sr+a^2$.
The perturbations around the Kerr background spacetime $\psi_s$ (a spin-$s$ field) can be obtained by solving the Teukolsky equation \cite{1973ApJ...185..635T}. Expanding the spin-$s$ field in spin-weighted spheroidal harmonics ${}_sS_{lm} (\theta)$,
\begin{equation} \displaystyle
\psi_{s} (t,r,\theta, \phi) = \frac{1}{2 \pi} \int e^{-i \omega t} \sum_{l=|s|}^{\infty} \sum_{m=-l}^{l} e^{im \phi} R_{lm} (r) {}_s S_{lm} (\theta) d\omega
\end{equation}
the Teukolsky equation reduces to the following form:
\begin{align}
&\Delta^{-s} \frac{d}{dr} \left( \Delta^{s+1} \frac{d R_{lm}}{dr} \right)- V_s R_{lm}=-T_s,
\label{eq:teuk1}\\
&V_s \equiv - \left(\frac{K^2-is K \Delta'}{\Delta}+2is K' -\lambda_s \right),
\end{align}
where $\lambda_s \equiv {}_sA_{lm} +a^2 \omega^2 -2 am\omega$, $K\equiv (r^2 + a^2) \omega -am$, and ${}_sA_{lm}$ is the separation constant. $T_s$ in (\ref{eq:teuk1}) is the source term and determines the excitation of QNMs. The spheroidal harmonics satisfies the following equation:
\begin{align}
\begin{split}
&\frac{1}{\sin\theta}\frac{d}{d\theta}\left(\sin{\theta} \frac{d {}_sS_{lm}}{d\theta}\right)\\
&+\left(a^2 \omega^2 \cos^2\theta-\frac{m^2}{\sin^2\theta}-2a\omega s \cos\theta-\frac{2ms\cos\theta}{\sin^2\theta}-s^2 \cot^2\theta+s+A_{slm}\right) {}_sS_{lm}=0,
\label{eq:teuk2}
\end{split}
\end{align}
and requiring the regularity of the solution (\ref{eq:teuk2}), one can obtain the value of ${}_s A_{lm}$ \cite{Berti:2005ys}.
The wave equation (\ref{eq:teuk1}) is imaginary and it has its long range potential. However, one can transform the Teukolsky equation by using the general Darboux transformation \cite{Chandrasekhar:1976zz}
\begin{equation}
{}_sX_{lm} = \Delta^{s/2} (r^2+a^2)^{1/2} \left( \xi (r) {}_sR_{lm} + \zeta (r) \Delta^{s+1} \frac{d {}_sR_{lm}}{dr} \right),
\end{equation}
where the two transformation functions $\xi$ and $\zeta$ are related by
\begin{equation}
\xi^2 -\xi' \zeta \Delta^{s+1} + \xi \zeta' \Delta^{s+1} - \zeta^2 \Delta^{2s +1} V_s = \text{constant.}
\label{cond_ab}
\end{equation}
Using the tortoise coordinate $d r^{\ast} \equiv dr (r^2+a^2) / \Delta$, one can obtain the following wave equation:
\begin{equation}
\frac{d^2 {}_sX_{lm}}{dr^{\ast} {}^2} - {\cal V} {}_sX_{lm} = - \tilde{T}_s,
\label{real_wave_eq1}
\end{equation}
where the potential is given by
\begin{align}
{\cal V} &\equiv \frac{\Delta U}{(r^2+a^2)^2} +G^2 + \frac{d G}{dr^{\ast}},\\
G &\equiv \frac{s (r-M)}{r^2+a^2} + \frac{r \Delta}{(r^2+a^2)^2},\\
U &\equiv V_s + \frac{2 \xi' + (\zeta' \Delta^{s+1})'}{\zeta \Delta^s},
\end{align}
and the explicit form of a transformed source term $\tilde{T}_s$ is given in \cite{Chandrasekhar:1976zz}. The choice of $\xi$ and $\zeta$ are arbitrary, provided that the condition (\ref{cond_ab}) is satisfied, and it is possible to choose the functions $\xi$ and $\zeta$ so that the resulting potential in the wave equation is purely real. The derivation of real angular momentum potentials was pioneered by Chandrasekhar and Detweiler \cite{Chandrasekhar:1976zz,Detweiler:1977gy}. They found four kinds of transformation functions
\begin{align}
\xi_{ij} &= \frac{2 \sqrt{2} \rho^4}{|\kappa| \Delta} \left[  \Xi_{i} + \Theta_{ij} \left( \frac{3 r \Delta}{\rho^4} +i \omega \right)\right],\\
\zeta_{ij} &= -2 \sqrt{2} \Delta \rho^2 \Theta_{ij}/|\kappa|,
\end{align}
where the suffixes $i$ and $j$ take $+1$ or $-1$ and the definition of the functions $\rho$, $\Xi_{ij}$, $\Theta_{ij}$, and $\kappa$ are shown in Appendix \ref{app:ex}. The resulting potential is given by
\begin{align}
\begin{split}
{\cal V}_{ij} &= \frac{-K^2}{(r^2+a^2)^2} + \frac{\rho^4 \Delta}{(r^2+a^2)^2} \left[ \frac{\lambda (\lambda+2)}{g+ b_i \Delta} -b_i \frac{\Delta}{\rho^8} + \frac{(\kappa_{ij} \rho^2 \Delta - h) (\kappa_{ij} \rho^2 g -b_i h)}{\rho^4 (g+b_i \Delta) (g-b_i \Delta)^2} \right]\\
&+ \left[ \frac{r \Delta a m/\omega}{(r^2+a^2)^2 \rho^2} \right]^2 - \frac{\Delta}{(r^2+a^2)} \frac{d}{dr} \left[ \frac{r \Delta a m/\omega}{(r^2+a^2)^2 \rho^2} \right].
\end{split}
\label{CD_potential_revised}
\end{align}
and
\begin{align}
b_{\pm1} &\equiv \pm 3 (a^2 - am/\omega),\\
\kappa_{ij} &\equiv j \left\{ 36 G^2 M^2 -2\lambda \left[ (a^2-am/\omega) (5\lambda+6) -12 a^2 \right] + 2 b_i \lambda (\lambda +2) \right\}^{1/2}.
\end{align}
Note that the real potentials ${\cal V}_{ij}$ become singular when $g(\omega)+ b_i (\omega) \Delta = 0$, but the singular frequency region is different for a different choice of ($i,j$). Therefore, we need to use the four real potentials complementarily to obtain regular mode functions in the entire frequency space. Recently, Glampedakis, Johnson, and Kennefick found out that solutions to the CD real wave equation with the potentials ${\cal V}_{-1, \pm1}$ and ${\cal V}_{+1, \mp1}$ are related by the Darboux transformation \cite{Glampedakis:2017rar}. The fact that the solutions with the different potentials are related by the Darboux transformation is important since it guarantees that they lead to the same reflection/transmission amplitudes and QNM spectra. We shall return to this point later to show the invariance of echo spectra under the generalized Darboux transformation.

\begin{figure}[t]
    \includegraphics[width=0.9\textwidth]{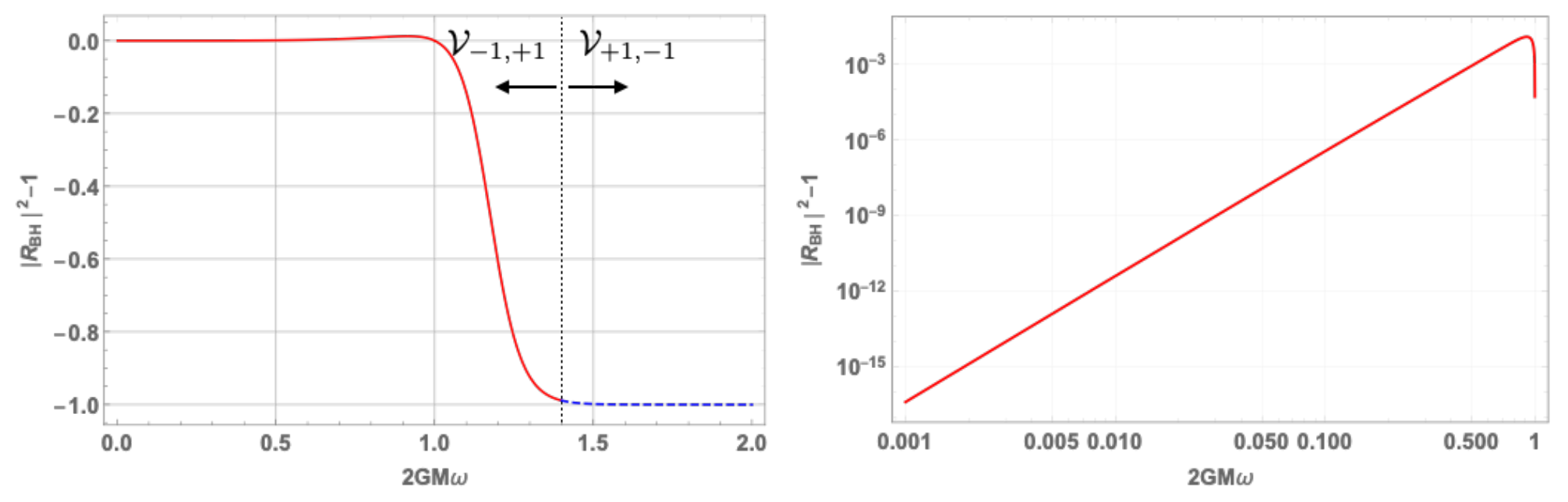}
\caption{The left panel shows the energy flux amplification factor $|{\cal R}_{\text{BH}}^{\rightarrow}|^2-1$ for $\bar{a} = 0.8$ and $l=m=2$. The red solid line is calculated by using the real potential of ${\cal V}_{-1,+1}$ and blue dashed line is calculated from ${\cal V}_{+1,-1}$ to avoid the singular behavior of ${\cal V}_{+1,-1}$ (${\cal V}_{-1,+1}$) at the low (high) frequency region. The right panel shows the same amplification factor for $\omega \leq m \Omega_H$ ($2GM\omega \leq 1$). The maximum amplification factor is around $0.0125$.
}
\label{det_reflection}
\end{figure}

In the following, we omit the subscripts $s$, $l$, and $m$ for brevity. The solution of a homogeneous (real) wave equation (\ref{real_wave_eq1}) has the asymptotic form of
\begin{align}
X \sim
\begin{cases}
A_{\text{in}} e^{-i \omega r^{\ast}} +A_{\text{out}} e^{i \omega r^{\ast}} \ \ \text{for} \ \ r^{\ast} \to \infty,\\
B_{\text{in}} e^{-i \tilde{\omega} r^{\ast}} +B_{\text{out}} e^{i \tilde{\omega} r^{\ast}} \ \ \text{for} \ \ r^{\ast} \to -\infty,
\end{cases}
\end{align}
where $A_{\rm in/out}$ and $B_{\rm in/out}$ represent the amplitudes of ingoing/outgoing mode functions at infinity and at horizon, respectively.
Since the complex conjugate of $X$ is also a solution of (\ref{real_wave_eq1}) due to the real potential, the Wronskian $W \equiv X dX^{\ast}/dr^{\ast} - X^{\ast} dX/dr^{\ast}$ is constant. Calculating the Wronskian at $r^{\ast} \to \pm \infty$ and using the Wronskian relation of $W(r^{\ast} = -\infty) = W(r^{\ast} = +\infty)$, one obtains the following relation:
\begin{equation}
\tilde{\omega} (|B_{\text{out}}|^2 - |B_{\text{in}}|^2)=
\omega (|A_{\text{out}}|^2 - |A_{\text{in}}|^2).
\end{equation}
If one considers the incident wave from outside/inside the potential, the above equation gives the energy conservation law for incident, reflected, and transmitted waves
\begin{align}
|A_{\text{in}}|^2 &= |A_{\text{out}}|^2 + \frac{\tilde{\omega}}{\omega} |B_{\text{in}}|^2,\\
|B_{\text{out}}|^2 &= |B_{\text{in}}|^2 + \frac{\omega}{\tilde{\omega}} |A_{\text{out}}|^2.
\end{align}
Then one obtains the the relation between the energy reflectivity and transmissivity
\begin{align}
1 &= |{\cal R}_{\text{BH}}^{\leftarrow}|^2 + \frac{\tilde{\omega}}{\omega} |{\cal T}_{\text{BH}}^{\leftarrow}|^2,\\
1 &= |{\cal R}_{\text{BH}}^{\rightarrow}|^2 + \frac{\omega}{\tilde{\omega}} |{\cal T}_{\text{BH}}^{\rightarrow}|^2,
\end{align}
where the amplitude reflectivities and transmissivities, ${\cal R}_{\text{BH}}^{\rightarrow}$ (${\cal R}_{\text{BH}}^{\leftarrow}$) and ${\cal T}_{\text{BH}}^{\rightarrow}$ (${\cal T}_{\text{BH}}^{\leftarrow}$), are defined by $B_{\text{in}}/B_{\text{out}}$ ($A_{\text{out}}/A_{\text{in}}$) and $A_{\text{out}}/B_{\text{out}}$ ($B_{\text{in}}/A_{\text{in}}$), respectively.
We show the numerically calculated energy flux amplification factor in FIG. \ref{det_reflection}. To cover the frequency region of $0.001 \leq 2GM \omega \leq 2$, we used two real potentials, ${\cal V}_{-1, +1}$ and ${\cal V}_{+1, -1}$, which are related by the Darboux transformation. 
Although it was shown that \cite{Glampedakis:2017rar} $\text{arg} [{\cal T}_{\text{BH}}^{\rightarrow}]$ is invariant under the Darboux transformation, $\text{arg} [{\cal R}_{\text{BH}}^{\rightarrow}]$ is not. Moreover, neither is invariant under the generalized Darboux transformation. However, in Sec. \ref{sec:phase_shift} we will show that ${\cal R}(\omega) {\cal R}_{\text{BH}}^{\rightarrow}$ is invariant under the transformation, which means that the frequency peaks in echo spectrum and the QNMs of GW echoes are invariant under the generalized Darboux transformation, as expected on physical grounds.

Although the CD potentials are real and short range (compared to Teukolsky), they are singular in a certain frequency region depending on the parameters $a$, $m$, and $\omega$. As another option, one can use the SN equation, which has a short range and regular potential for any parameters and in the whole frequency space, but it has a generally complex potential. Sasaki and Nakamura have transformed Eq. (\ref{eq:teuk1}) so that it reduces to the Regge-Wheeler equation in the limit of $a \to 0$ and it has the form of \cite{10.1143/PTP.67.1788}
\begin{equation}
\left( \frac{d^2}{dr^{\ast} {}^2} -{}_sF_{lm}(r) \frac{d}{dr^{\ast}}- {}_sU_{lm} (r) \right) X = -T_{SN},
\label{SNeq}
\end{equation}
where $T_{SN}$ is the source term in the SN expression, and the explicit forms of ${}_sF_{lm}$ and ${}_sU_{lm}$ are given in \cite{10.1143/PTP.67.1788,Tagoshi:1996gh}. The SN equation (\ref{SNeq}) asymptotically reduces to
\begin{align}
\begin{cases}\displaystyle
\left( \frac{d^2}{dr^{\ast} {}^2} + \tilde{\omega}^2 \right) X = - T_{SN}, \ &\text{for} \ r^{\ast} \to - \infty,\\
\displaystyle
\left( \frac{d^2}{dr^{\ast} {}^2} + \omega^2 \right) X = -T_{SN}, \ &\text{for} \ r^{\ast} \to + \infty.
\end{cases}
\label{asymSNeq}
\end{align}
Therefore, the homogeneous SN equation has the solution of superposition of ingoing and outgoing modes at the asymptotic regions $r^{\ast} \to \pm \infty$
\begin{equation}
{}_s X_{lm} =
\begin{cases}
A e^{i \tilde{\omega} r^{\ast}} + B e^{-i \tilde{\omega} r^{\ast}} & \text{for} \ r^{\ast} \to - \infty,\\
C e^{i \omega r^{\ast}} + D e^{-i \omega r^{\ast}} & \text{for} \ r^{\ast} \to  +\infty,
\end{cases}
\end{equation}
where $A$, $B$, $C$, and $D$ are arbitrary constants.

In order to solve the homogeneous wave equations of (\ref{real_wave_eq1}) or (\ref{SNeq}), imposing a proper boundary condition is necessary and the boundary condition should be determined by the physical property of a spinning BH's horizon. Although only the ingoing mode $e^{-i \tilde{\omega} r^{\ast}}$ is allowed in the near-horizon limit in general relativity, it could be also true that the quantum gravitational effect may lead to a nontrivial configuration at $r\simeq r_+ + \delta r$ with $\delta r/r_+ \ll 1$, which partially reflects ingoing GWs near the surface.
Imposing a boundary condition of reflectivity ${\mathcal R} (\omega)$, the mode function satisfies
\begin{align}
X \sim
\begin{cases}
e^{-i \tilde{\omega} r^{\ast}} + {\mathcal R} (\omega) e^{i \tilde{\omega} r^{\ast}} \ &\text{for} \ r^{\ast} \to - \infty,\\
e^{i \omega r^{\ast}} \ &\text{for} \ r^{\ast} \to + \infty.
\label{ref_boundary}
\end{cases}
\end{align}
The constant and Boltzmann reflectivity models give ${\cal R} ( \omega) = R_c e^{i\delta_{\text{wall}}}$ and ${\cal R} ( \omega) = e^{-|\tilde{\omega}|/(2T_H)} e^{i\delta_{\text{wall}}}$, respectively.
Since the Hawking temperature approaches to zero temperature in the extremal limit, the Boltzmann reflectivity of a rapidly spinning quantum BH may be exponentially suppressed.

\subsection{Green's function technique and transfer function for GW echoes}
In this subsection, we briefly review the technique to obtain the solution of the wave equation with a reflective boundary. Let us begin with the calculation of the Green's function based on the CD equation
\begin{equation}
\left( \frac{d^2}{dr^{\ast} {}^2} - {\cal V}_{ij} \right) G(r^{\ast}-r^{\ast} {}') = \delta (r^{\ast}-r^{\ast} {}').
\end{equation}
In order to construct the Green's function with the reflective boundary, we need two linear independent homogeneous solutions:
\begin{align}
&X^{(\text{in})} \sim
\begin{cases}
e^{-i \tilde{\omega} r^{\ast}} \ &\text{for} \ r^{\ast} \to - \infty,\\
A_{\text{out}} e^{i \omega r^{\ast}} + A_{\text{in}} e^{-i \omega r^{\ast}} \ &\text{for} \ r^{\ast} \to + \infty,
\label{modein}
\end{cases}\\
&X^{(\text{up})} \sim
\begin{cases}
B_{\text{in}} e^{-i \tilde{\omega} r^{\ast}} + B_{\text{out}} e^{i \tilde{\omega} r^{\ast}} \ &\text{for} \ r^{\ast} \to - \infty,\\
e^{i \omega r^{\ast}} \ &\text{for} \ r^{\ast} \to + \infty,
\label{modeout}
\end{cases}
\end{align}
where $X^{\rm(in)}$ and $X^{\rm(up)}$ represent the homogeneous solutions with an ingoing mode at horizon and with an outgoing mode at infinity, respectively.
Using the mode functions (\ref{modein}) and (\ref{modeout}), one can construct the Green's function for the boundary condition (\ref{ref_boundary}) with ${\mathcal R} (\omega) = 0$, which is given by
\begin{equation}
G_{\text{BH}} (r^{\ast}, r^{\ast} {}') = \frac{X^{(\text{in})} (r^{\ast}_{<}) X^{(\text{up})} (r^{\ast}_>)}{W_{\text{BH}}},
\end{equation}
where $r^{\ast}_{<} \equiv \text{min} (r^{\ast}, r^{\ast} {}')$, $r^{\ast}_{>} \equiv \text{max} (r^{\ast}, r^{\ast} {}')$, and $W_{\text{BH}} \equiv 2i \tilde{\omega} B_{\text{out}}$ is the Wronskian of $X^{(\text{in})}$ and $X^{(\text{up})}$. Therefore, the Fourier mode of GWs at infinity and near the horizon for ${\mathcal R} = 0$ can be obtained as
\begin{align}
\lim_{r^{\ast} \to \infty} X(r^{\ast}) &= -X^{(\text{up})} (r^{\ast}) \int^{+ \infty}_{-\infty} d r^{\ast} {}' \frac{X^{(\text{in})} (r^{\ast} {}') \tilde{T} (r^{\ast} {}')}{W_{\text{BH}}} \equiv X^{(\text{up})} Z_{\infty} (\omega),
\label{Z_inf}\\
\lim_{r^{\ast} \to -\infty} X(r^{\ast}) &= -X^{(\text{in})} (r^{\ast}) \int^{+ \infty}_{-\infty} d r^{\ast} {}' \frac{X^{(\text{up})} (r^{\ast} {}') \tilde{T} (r^{\ast} {}')}{W_{\text{BH}}} \equiv X^{(\text{in})} Z_{\text{BH}} (\omega).
\label{Z_BH}
\end{align}
When ${\mathcal R} \neq 0$, the Schwarzschild Green's function, $G_{\text{BH}}$, needs a modification since the ingoing GWs near the horizon, whose amplitude is $Z_{\text{BH}}$, also contribute to the GWs at infinity. Using the geometric optics approximation \cite{Mark:2017dnq}, one can obtain the amplitude of the $n$ th echo 
\begin{equation}
Z^{(n)}_{\text{BH}} = {\mathcal T}_{\text{BH}} {\mathcal R}^{n} {\mathcal R}_{\text{BH}} {}^{n-1} e^{-2in\tilde{\omega} x_0} \times Z_{\text{BH}} (\omega),
\label{geodesic_series1}
\end{equation}
and the total echo amplitude is given by
\begin{equation}\displaystyle
\sum_{n=1}^{\infty} Z_{\text{BH}}^{(n)} = \frac{{\mathcal T}_{\text{BH}} {\mathcal R} e^{-2i \tilde{\omega} x_0}}{1-{\mathcal R} {\mathcal R}_{\text{BH}} e^{-2i \tilde{\omega} x_0}} Z_{\text{BH}} \equiv {\mathcal K} (\omega) Z_{\text{BH}}.
\end{equation}
Note that the transfer function ${\cal K} (\omega)$ is derived from the sum of the geometric series of the common ratio of ${\cal R} {\cal R}_{\text{BH}}$. If $|{\cal R} {\cal R}_{\text{BH}}|\geq 1$, the sum of the geometric series does not converge, i.e., the Fourier transform of an unstable system is not well defined. Finally, we obtain the spectrum of echoes for a spinning BH based on the geometric optics approximation as follows:
\begin{equation}
X (r^{\ast}) = -\int^{+\infty}_{-\infty} dr^{\ast} {}' G(r^{\ast}, r^{\ast} {}') \tilde{T} (r^{\ast} {}') = (Z_{\infty} +{\mathcal K} Z_{\text{BH}}) e^{i \omega r^{\ast}} \equiv \tilde{X} e^{i\omega r^{\ast}}.
\label{echo_spectrum}
\end{equation}
Implementing the inverse Fourier transform of (\ref{echo_spectrum}), one can obtain the corresponding time domain function.

\subsection{Superradiance and ergoregion instability}
\begin{figure}[t]
    \includegraphics[width=0.7\textwidth]{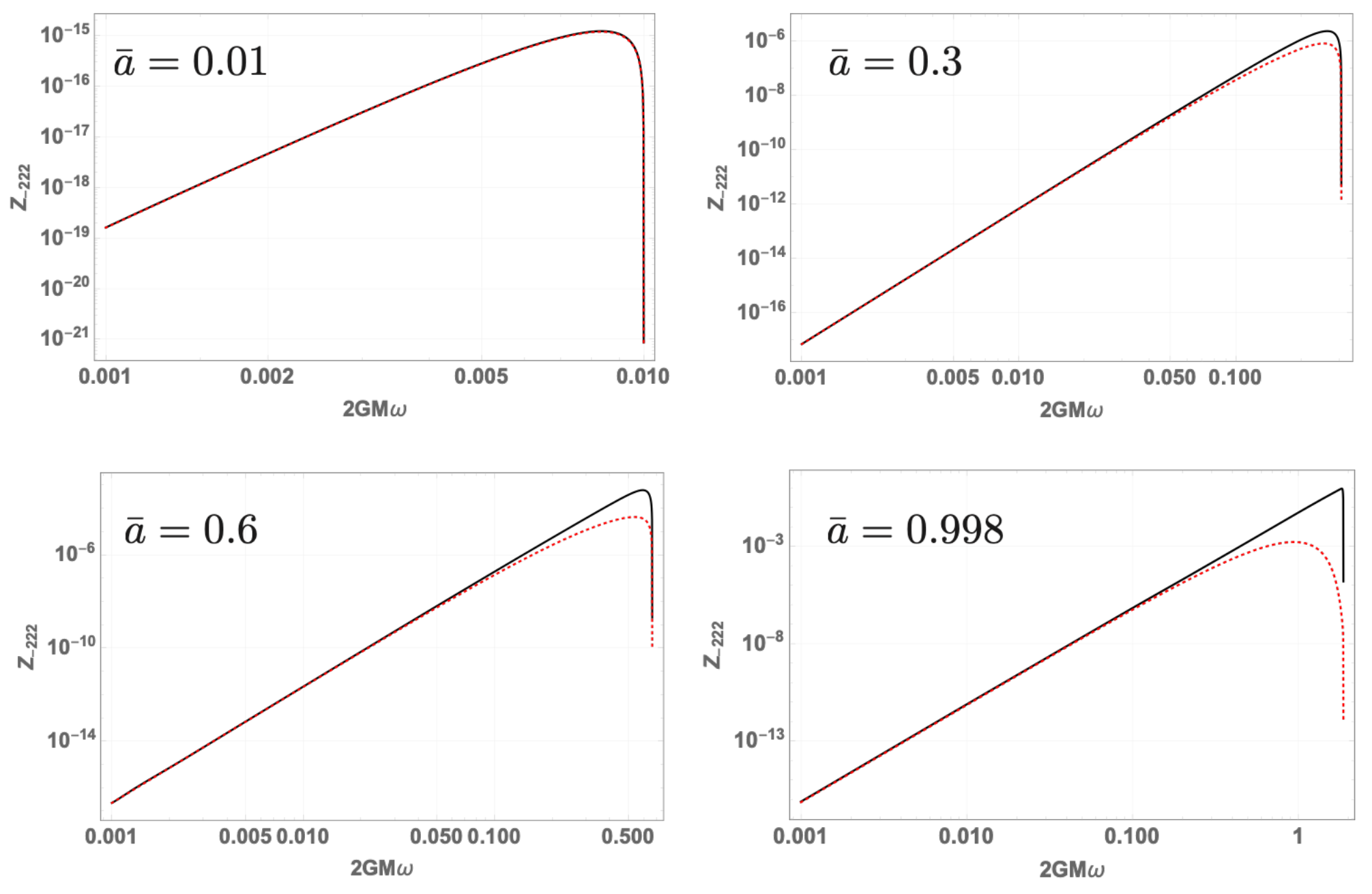}
\caption{The frequency dependence of the amplification factors with $\bar{a}=0.01, 0.3, 0.6,$ and $0.998$ are shown (black solid lines). Red dashed lines are the analytic function of (\ref{sta-san}).
}
\label{superrad1}
\end{figure}

The reflection near the BH horizon and superradiance may lead to the ergoregion instability, which gives the constraint on the reflectivity of a spinning BH. The superradiance is a process by which incoming radiation is enhanced and reflected by extracting the rotation energy of the spinning BH. The amplification factor is a good measure of superradiance.
In the CD expression, the amplification factor of a spinning BH, $Z_{slm}$, is
\begin{equation}
Z_{slm} \equiv \frac{|B_{\text{in}}|^2}{|B_{\text{out}}|^2} -1 = |{\cal R}_{\text{BH}}^{\rightarrow}|^2-1.
\end{equation}
We only consider the reflection/transmission of outgoing incident waves in the following, and so we will omit the symbol ``$\rightarrow$". We calculate the reflection and transmission coefficients, ${\cal R}_{\text{BH}}$ and ${\cal T}_{\text{BH}}$, by numerically integrating the CD equation and the obtained amplification factor, $Z_{-222} = |{\cal R}_{\text{BH}}|^2-1$, is shown in FIG. \ref{superrad1}. We also calculate the amplification factor for various spin parameters by using the SN equation to check the consistency (with details provided in Appendix \ref{app:SN}).
These numerical results are consistent with the analytic expression of the amplification factor for the low-frequency regime, $\omega GM \ll 1$ \cite{1973JETP...37...28S}:\footnote{For a spinning BH resulting from a $30M_\odot-30M_\odot$ binary BH merger, the analytic expression in (\ref{sta-san}) may be a good approximation up to a few $10$ Hz, but deviates from the numerical result from $\sim 100$ Hz.}
\begin{equation}\displaystyle
Z_{slm} \simeq 4 Q \beta_{sl} \prod_{k=1}^{l} \left( 1+ \frac{4 Q^2}{k^2} \right) [\omega (r_+ -r_-)]^{2l +1},
\label{sta-san}
\end{equation}
where $r_-$ is the radius of the inner horizon, $\sqrt{\beta_{sl}} \equiv \frac{(l-s)! (l+s)!}{(2l)! (2l+1)!!}$ and $Q \equiv- \frac{r_+^2 +a^2}{r_+-r_-} \tilde{\omega}$. We numerically obtain the amplification factor of $Z_{-222} \simeq 0.91$ for $\bar{a}=0.998$ which is the maximum spin known as the Thorne limit.

Remember that $|{\cal R} {\cal R}_{\text{BH}}| \geq 1$ leads to the divergence of the infinite sum of the geometric series in (\ref{geodesic_series1}), which is nothing but the ergoregion instability \cite{PhysRevD.99.064007}. One can understand this from the point of view of the QNMs of a reflective BH. The poles in the Green's function of GW echoes can be obtained by looking for the zero points of the denominator of the transfer function ${\cal K}$
\begin{equation}
1-{\cal R} {\cal R}_{\text{BH}} e^{-2i \tilde{\omega}_n x_0} = 0.
\end{equation}
Solving this equation in terms of the QNMs $\tilde{\omega}_n$, one obtains
\begin{equation}
\omega_n = \frac{2 \pi n +(\delta_{\text{wall}}+ \delta_{\text{BH}})}{\Delta t_{\text{echo}}} + m \Omega_H + i \frac{\ln{|{\cal R} {\cal R}_{\text{BH}}|}}{\Delta t_{\text{echo}}},
\end{equation}
where $\Delta t_{\text{echo}} \equiv 2 |x_0|$, $\delta_{\text{wall}} \equiv \text{arg} [{\cal R}]$ and $\delta_{\text{BH}} \equiv \text{arg} [{\cal R}_{\text{BH}}]$. Then we obtain the real and imaginary parts of the echo QNMs
\begin{align}
\text{Re} [\omega_n] &\simeq \frac{2 \pi n +(\delta_{\text{wall}}+ \delta_{\text{BH}})}{\Delta t_{\text{echo}}} + m \Omega_H,\label{re_omega}\\
\text{Im} [\omega_n] &\simeq \left. \frac{\ln{|{\cal R} {\cal R}_{\text{BH}}|}}{\Delta t_{\text{echo}}} \right|_{\omega = \text{Re}[\omega_n]},
\end{align}
and in the low-frequency regime $(GM \omega \ll 1)$, one can obtain the analytic expression of the imaginary part of QNMs by using (\ref{sta-san})
\begin{equation}
\text{Im} [\omega_n] \simeq \frac{\ln|{\cal R}|}{\Delta t_{\text{echo}}} + \frac{2Q}{\Delta t_{\text{echo}}} \beta_{sl} \prod_{k=1}^{l} \left( 1+ \frac{4 Q^2}{k^2} \right) \left[\left(  \frac{2 \pi n +(\delta_{\text{wall}}+ \delta_{\text{BH}})}{\Delta t_{\text{echo}}} + m \Omega_H \right) (r_+ -r_-) \right]^{2l +1}.
\end{equation}
Therefore, the positivity of the imaginary part of QNMs, which leads to the instability, is equivalent to having $|{\cal R} {\cal R}_{\text{BH}}| > 1$. One can also see that the real part of QNM frequency only depends on the phases of the reflectivities at the would-be horizon and the angular momentum barrier.

\begin{figure}[t]
    \includegraphics[width=0.8\textwidth]{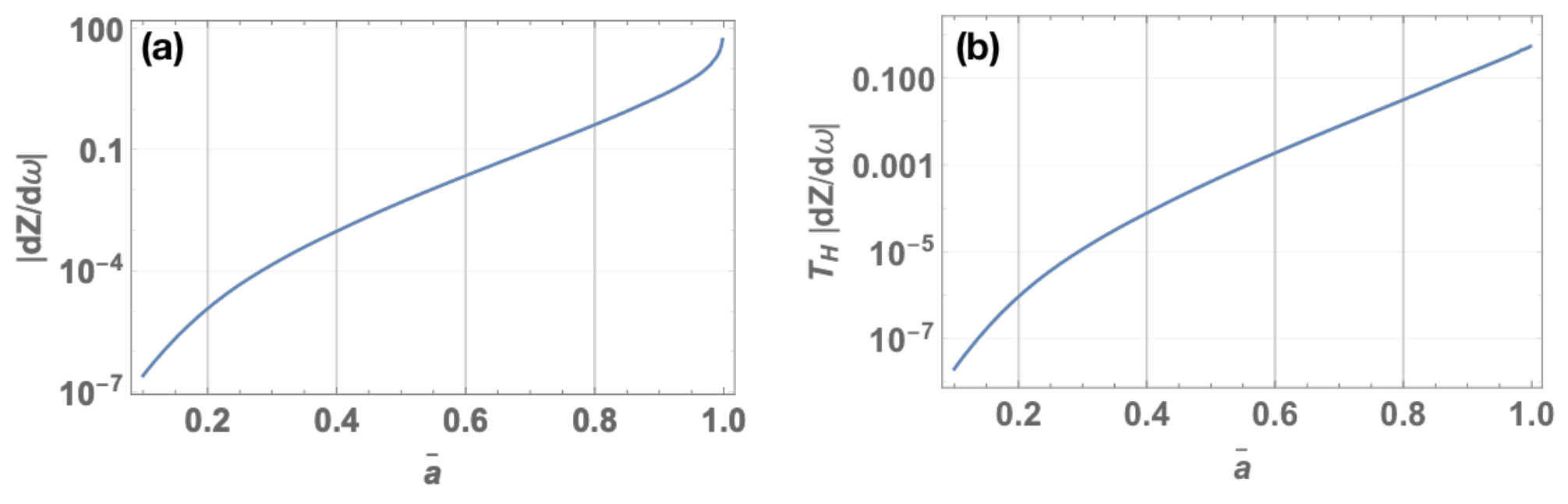}
\caption{Plots of (a) $|dZ/d\omega|_{\omega = m \Omega_H}$ and (b) $T_H |dZ/d\omega|_{\omega = m \Omega_H}$ in the spin range of $0.1 \leq \bar{a} \leq 0.998$.
}
\label{slope_fig}
\end{figure}

\begin{figure}[h]
    \includegraphics[width=0.5\textwidth]{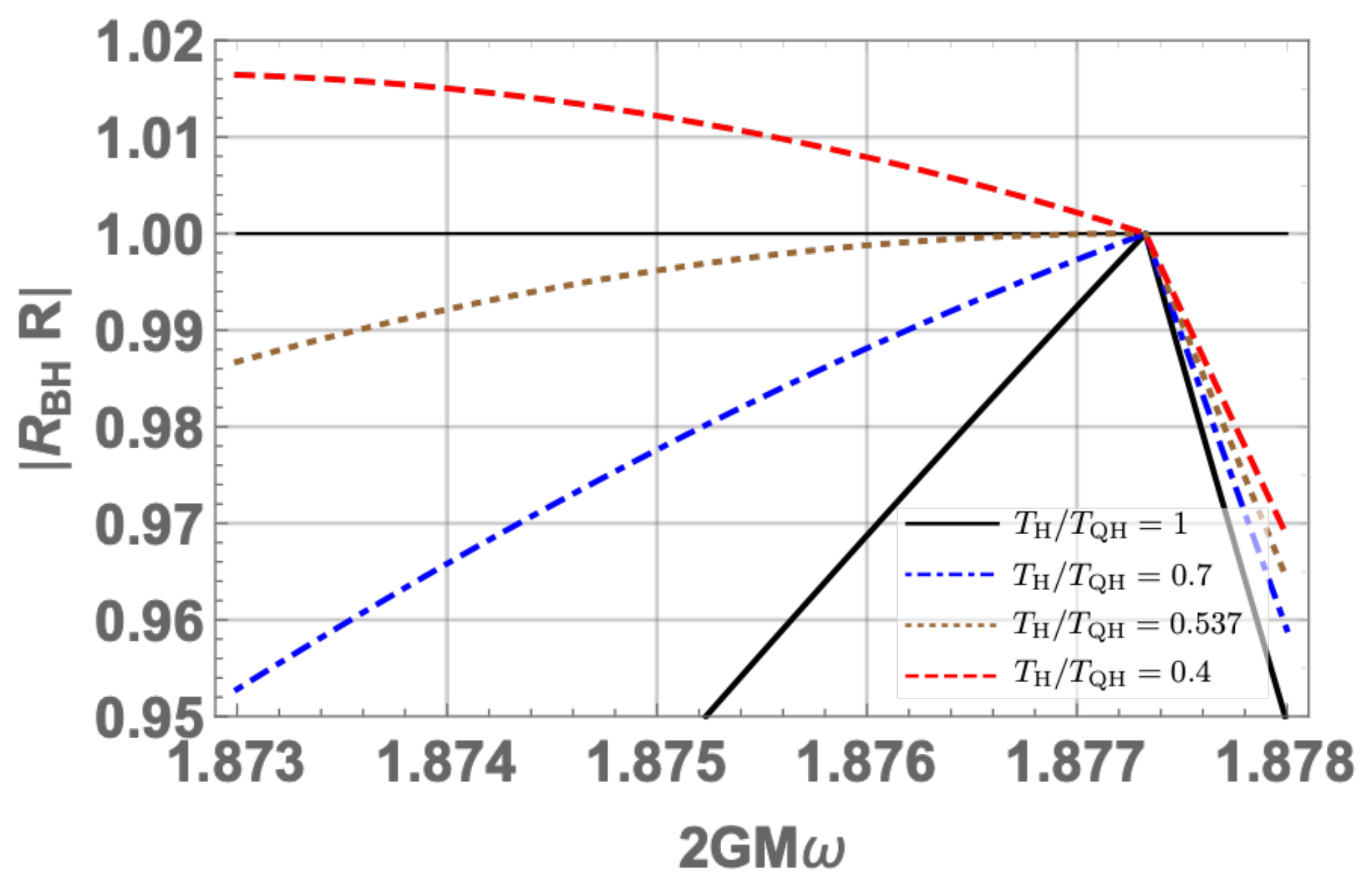}
\caption{Plot of $|{\cal R} {\cal R}_{\text{BH}}|$ with the Boltzmann reflectivity model. The spin parameter is $\bar{a} = 0.998$.
}
\label{superrad2}
\end{figure}

Using the criterion of $|{\cal R} {\cal R}_{\text{BH}}| < 1$, which guarantees that there is no ergoregion instability, we put the constraints on $R_c$ (FIG. \ref{C_super}) and $T_{\rm H} / T_{\rm QH}$ (FIG. \ref{B_super}) up to the Thorne limit \cite{Thorne:1974ve}. The parameters characterizing the reflectivity such that $|{\cal R} {\cal R}_{\text{BH}}|$ exceeds unity should be excluded. The Boltzmann reflectivity has its perfect reflectivity at $\omega = m \Omega_H$ and it can be approximated as 
\begin{equation}
|{\cal R}| = 1-\frac{1}{2 T_{\rm QH}} |\tilde{\omega}| \ \ \text{for} \ \ \omega \simeq m \Omega_H.
\end{equation}
On the other hand the amplification factor is approximated as $Z \simeq (dZ/d \omega) \tilde{\omega} \ll 1$ near the horizon frequency. In order for the ergoregion instability to be suppressed, the following inequality should be satisfied
\begin{equation}
|{\cal R} {\cal R}_{\text{BH}}| = e^{- |\tilde{\omega}|/(2T_{\rm QH})} \sqrt{1+Z} \simeq 1+ \left( \frac{1}{2T_{\rm QH}} + \frac{1}{2} \frac{dZ}{d\omega} \right) \tilde{\omega} \leq 1 \ \ \text{for} \ \ \tilde{\omega} \leq 0,
\end{equation}
from which one can obtain the lower bound for $T_{\rm H} / T_{\rm QH}$ as
\begin{equation}
T_{\rm H} / T_{\rm QH} \geq -T_{\rm H} (dZ/d\omega)|_{\omega = m \Omega_H}.
\end{equation}
In FIG. \ref{slope_fig}, we show the values of $|dZ/d\omega|$ at $\omega = m \Omega_H$ up to $\bar{a} \leq 0.998$, and one finds the lower bound of $T_{\rm H} / T_{\rm QH} \gtrsim 0.537$. In FIG. \ref{superrad2} we also plot $|{\cal R} {\cal R}_{\text{BH}}|$ for $\bar{a} =0.998$ and one can see $T_{\rm H} / T_{\rm QH} \gtrsim 0.537$ suppresses the ergoregion instability. For the constant reflectivity, we have the maximum amplification factor of $0.91$ at $\bar{a} =0.998$ that gives $R_c \lesssim 0.72$.

\subsection{Modeling the initial data}
\begin{figure}[b]
    \includegraphics[width=0.5\textwidth]{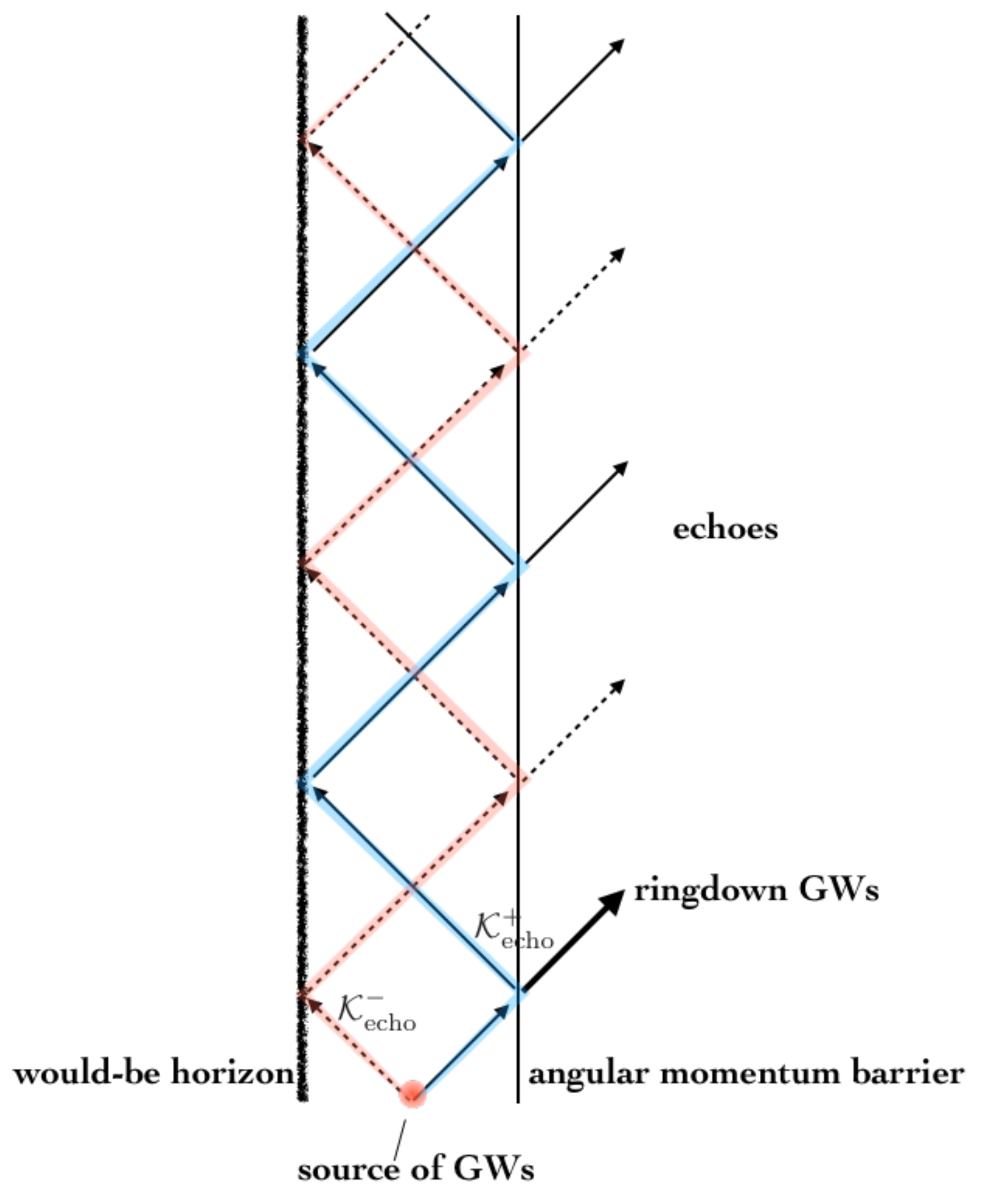}
\caption{A schematic picture showing the reflections of GWs sourced by $\tilde{T} = C(\omega) \delta (r^{\ast} - x_s)$ at the angular momentum barrier and would-be horizon.
}
\label{echo_delta_func}
\end{figure}

When the source term is located at $r^{\ast} = x_s$, that is $\tilde{T} = C(\omega) \delta (r^{\ast} - x_s)$, Eqs. (\ref{Z_inf}) and (\ref{Z_BH}) give a relation between $Z_{\infty}$ and $Z_{\text{BH}}$
\begin{equation}
\frac{Z_{\text{BH}}}{Z_{\infty}} = \frac{{\mathcal R}_{\text{BH}} + e^{-2i\tilde{\omega} x_s}}{{\mathcal T}_{\text{BH}}},
\end{equation}
which is independent of the function $C(\omega)$ \cite{PhysRevD.98.044018}.
Plugging this into (\ref{echo_spectrum}), one can read
\begin{align}
\tilde{X} &= Z_{\infty} \left( 1+ {\mathcal K} \frac{Z_{\text{BH}}}{Z_{\infty}} \right) =
Z_{\infty} \left( 1+ {\mathcal K}^{+}_{\text{echo}} + {\mathcal K}^{-}_{\text{echo}} \right),
\label{Xtilde}\\
{\mathcal K}^{+}_{\text{echo}} &\equiv \frac{{\mathcal R}_{\text{BH}} {\mathcal R} e^{-2i \tilde{\omega} x_0}}{1-{\mathcal R} {\mathcal R}_{\text{BH}} e^{-2i \tilde{\omega} x_0}},
\label{K+}\\
{\mathcal K}^{-}_{\text{echo}} &\equiv \frac{e^{-2i\tilde{\omega} x_s} {\mathcal R} e^{-2i \tilde{\omega} x_0}}{1-{\mathcal R} {\mathcal R}_{\text{BH}} e^{-2i \tilde{\omega} x_0}}.
\label{K-}
\end{align}
Based on the geometric optics picture, we can provide a reasonable interpretation for ${\mathcal K}^{+}_{\text{echo}}$ and ${\mathcal K}^{-}_{\text{echo}}$. Let us expand (\ref{K+}) and (\ref{K-}) in the form of the summations of geometric sequences
\begin{align}
{\mathcal K}^{+}_{\text{echo}} &= \sum_{n=1}^{\infty} {\mathcal R}^n {\mathcal R}_{\text{BH}}^n e^{-2i \tilde{\omega} x_0},
\label{K+se}\\
{\mathcal K}^{-}_{\text{echo}} &= e^{-2i \tilde{\omega} x_s} \sum_{n=1}^{\infty} {\mathcal R}^n {\mathcal R}_{\text{BH}}^{n-1} e^{-2i \tilde{\omega} x_0}.
\label{K-se}
\end{align}
One finds that $n$ reflections are involved in the $n$ th echo in (\ref{K+se}), and on the other hand, $(n-1)$ reflections and phase shift of $2 x_s \times \tilde{\omega}$ are involved in (\ref{K-se}). This implies that we now have two echo trajectories between the would-be horizon and the angular momentum barrier (FIG. \ref{echo_delta_func}).
Since we are interested in GW echoes connected to the BH ringdown, we will omit ${\mathcal K}^-_{\text{echo}}$ in the following, and will use the transfer function of
\begin{equation}
\tilde{X}/Z_{\infty} = 1+ {\cal K}_{\text{echo}}^{+} = \frac{1}{1-{\mathcal R} {\mathcal R}_{\text{BH}} e^{-2i \tilde{\omega} x_0}},
\end{equation}
which is equivalent to the transfer function derived in \cite{Wang:2019rcf}.

\begin{figure}[h]
\includegraphics[width=0.99\textwidth]{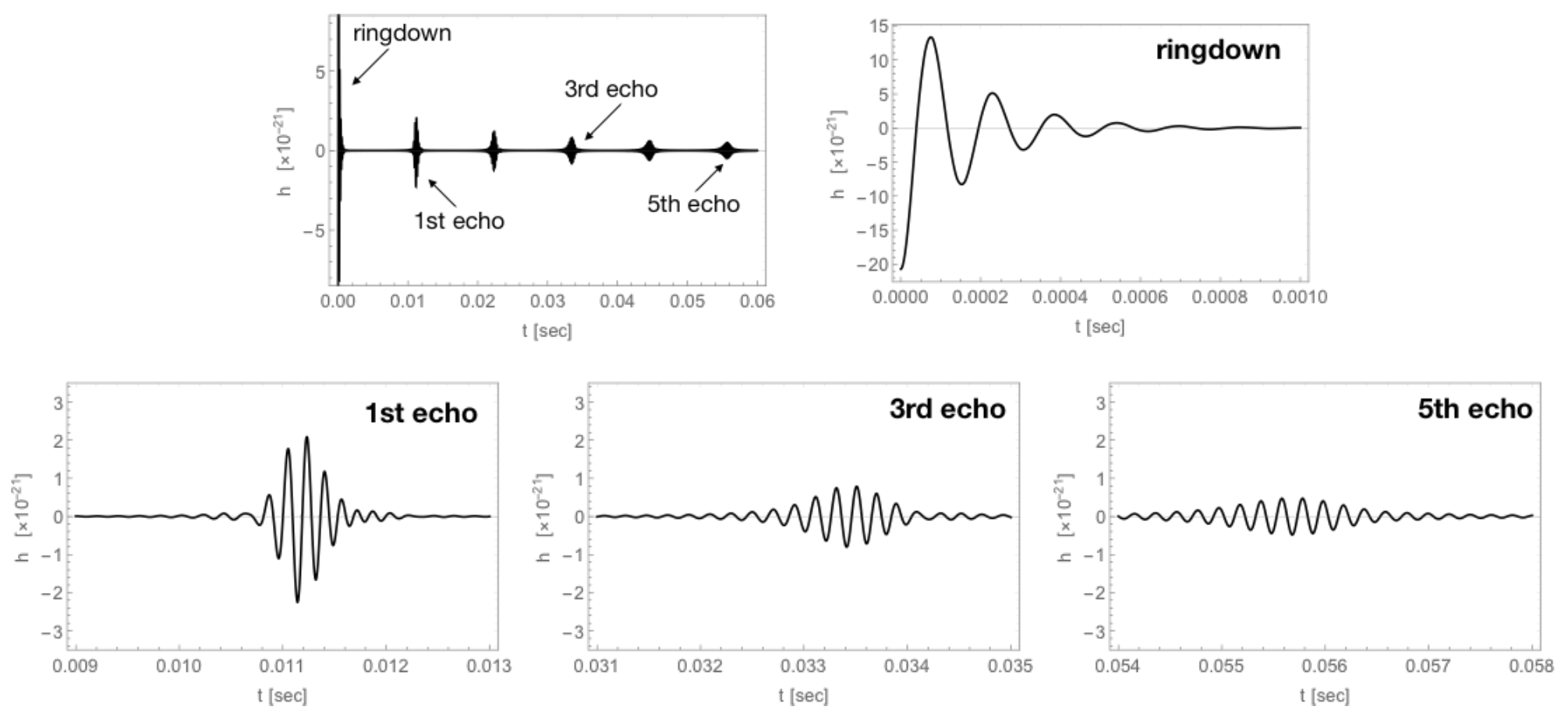}
\caption{A time domain function with $M=2.7 M_{\odot}$, $\bar{a} = 0.7$, $D_o=40$ Mpc, $\epsilon_{\text{rd}} =0.04$, $\theta=90^\circ$, and $\ell =m = 2$ in the Boltzmann reflectivity model with $\gamma=1$, $T_{\rm H} / T_{\rm QH} = 0.6$, and $x_0=x_{\text{m}}$.
}
\label{time_domain}
\end{figure}

We assume that GW echoes are excited by GW ringdown, whose spectrum is well expressed by the QNMs $\omega_{lmn}$ $(n=0,1,2,...)$. We model the amplitude of the ringdown signal $Z_{\infty}$ as \cite{Berti:2005ys}
\begin{align}
Z_{\infty} &= \frac{2 G M}{D_o} \tilde{A}_{lmn} \left[ e^{i \phi_{lmn}} S_{lmn} (\theta) \alpha_+ + e^{-i \phi_{lmn}} S^{\ast}_{lmn} (\theta) \alpha_- \right],
\label{Z_inf_QNM}\\
\alpha_{\pm} &\equiv \frac{-\text{Im}[\omega_{lmn}]}{\text{Im}[\omega_{lmn}]^2 + (\omega \pm \omega_{lmn})^2},
\end{align}
where $D_o$ is the distance between the GW source and observer, $\omega_{lmn}$ is the $n$ th QNM ($\omega_{lm0}$ is the least damping QNM), $\tilde{A}_{lmn}$ is the initial amplitude, $\theta$ is the viewing angle, and $\phi_{lmn}$ is the phase of ringdown GWs.
The amplitude of initial data $\tilde{A}_{lmn}$ is related to the radiation efficiency, $\epsilon_{\text{rd}}$, defined as
\begin{equation}
\epsilon_{\text{rd}} \equiv \frac{E_{\text{GW}}}{M},
\end{equation}
where $E_{\text{GW}}$ is the energy of ringdown GWs emitted in the $(l,m,n)$ QNM. Then the relation between $\tilde{A}_{lmn}$ and $\epsilon_{\text{rd}}$ is given by \cite{Berti:2005ys}
\begin{equation}
\tilde{A}_{lmn} = \sqrt{\frac{16 \pi^2 \epsilon_{\rm rd}}{M \text{Im}[\omega_{lmn}]^2 F}} \simeq \sqrt{\frac{16 \pi \epsilon_{\text{rd}}}{GM Q_{lmn} \omega_{lmn}}},
\label{eq:QNM_amplitude}
\end{equation}
where $Q_{lmn}$ is the quality factor of QNM of $\omega_{lmn}$, whose fitting function is provided in \cite{Berti:2005ys}, and $F$ is
\begin{equation}
F \equiv \int_0^{\infty} d\omega \omega^2 \left( \frac{1}{((\omega+\text{Re}[\omega_{lmn}])^2 + \text{Im}[\omega_{lmn}]^2)^2} + \frac{1}{((\omega-\text{Re}[\omega_{lmn}])^2 + \text{Im}[\omega_{lmn}]^2)^2} \right).
\end{equation}
The second near equality in (\ref{eq:QNM_amplitude}) is a good approximation for $n=0$. In the following, we will use $n=0$, the least damping QNM, to model $Z_{\infty}$, but we will show that the overtones ($n> 0$) can enhance the amplitude of GW echoes in the low-frequency region in Sec. \ref{sec:overtone}.
In FIG. \ref{time_domain}, as an example, we show a time domain function of ringdown and echo phases by implementing the inverse Fourier transform of $\tilde{X} (\omega)$:
\begin{equation}
h (t, r^{\ast}) \equiv \left|\int_{- \infty}^{\infty} d \omega \tilde{X}(\omega) e^{i \omega (r^{\ast} - t)} \right|.
\end{equation}

\section{Echo spectrum}
\label{sec:echo_spectrum}

\subsection{ECOs or the Planckian corrections in the dispersion relation?}
\label{sec:x_0}
\begin{figure}[h]
\includegraphics[width=0.45\textwidth]{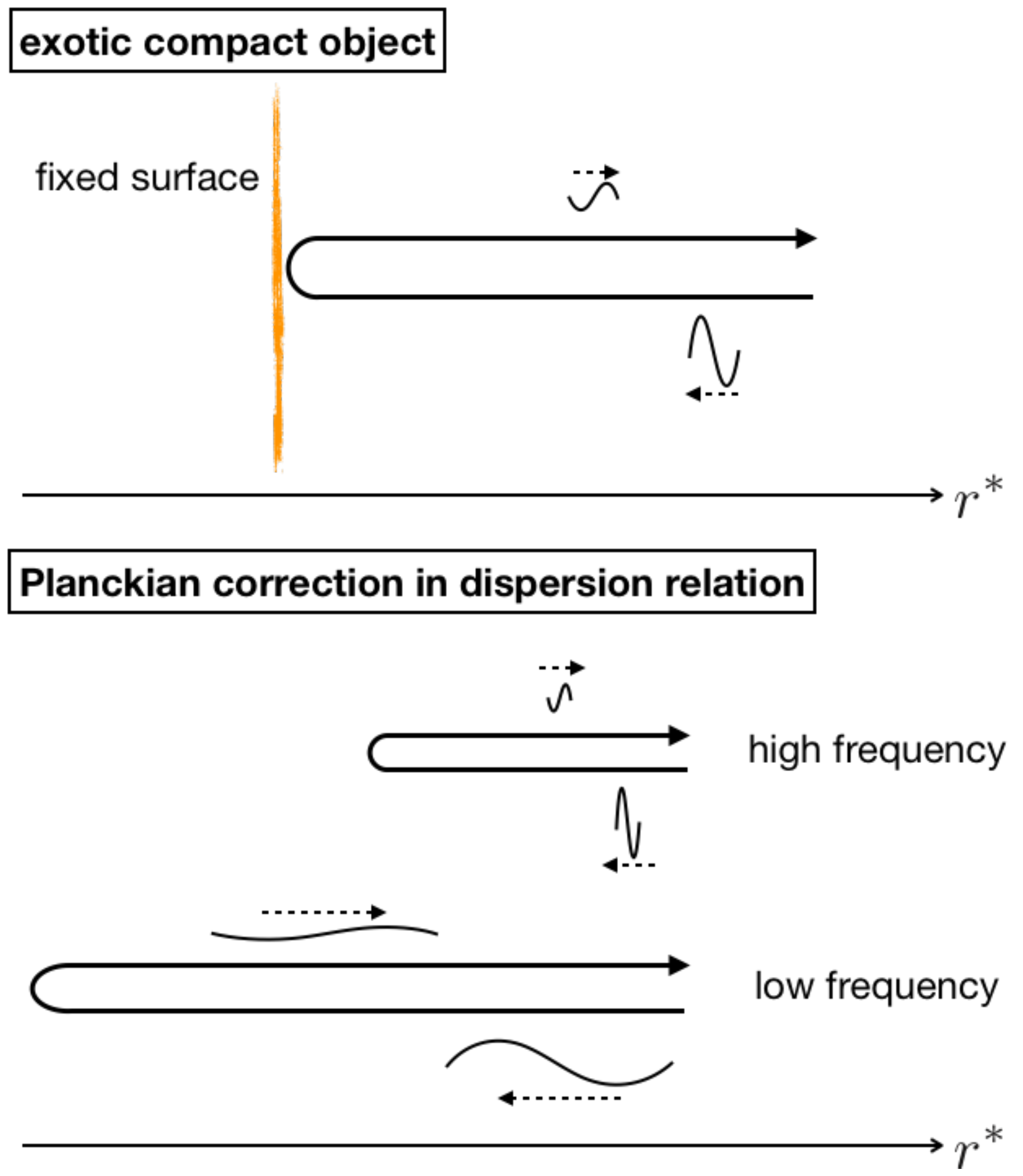}
\caption{Schematic picture showing two echo mechanisms based on the reflection due to exotic compact objects (top) and the reflection due to the Planckian correction in the dispersion relation (bottom).
}
\label{exo_pla}
\end{figure}
\begin{figure}[t]
\includegraphics[width=0.8\textwidth]{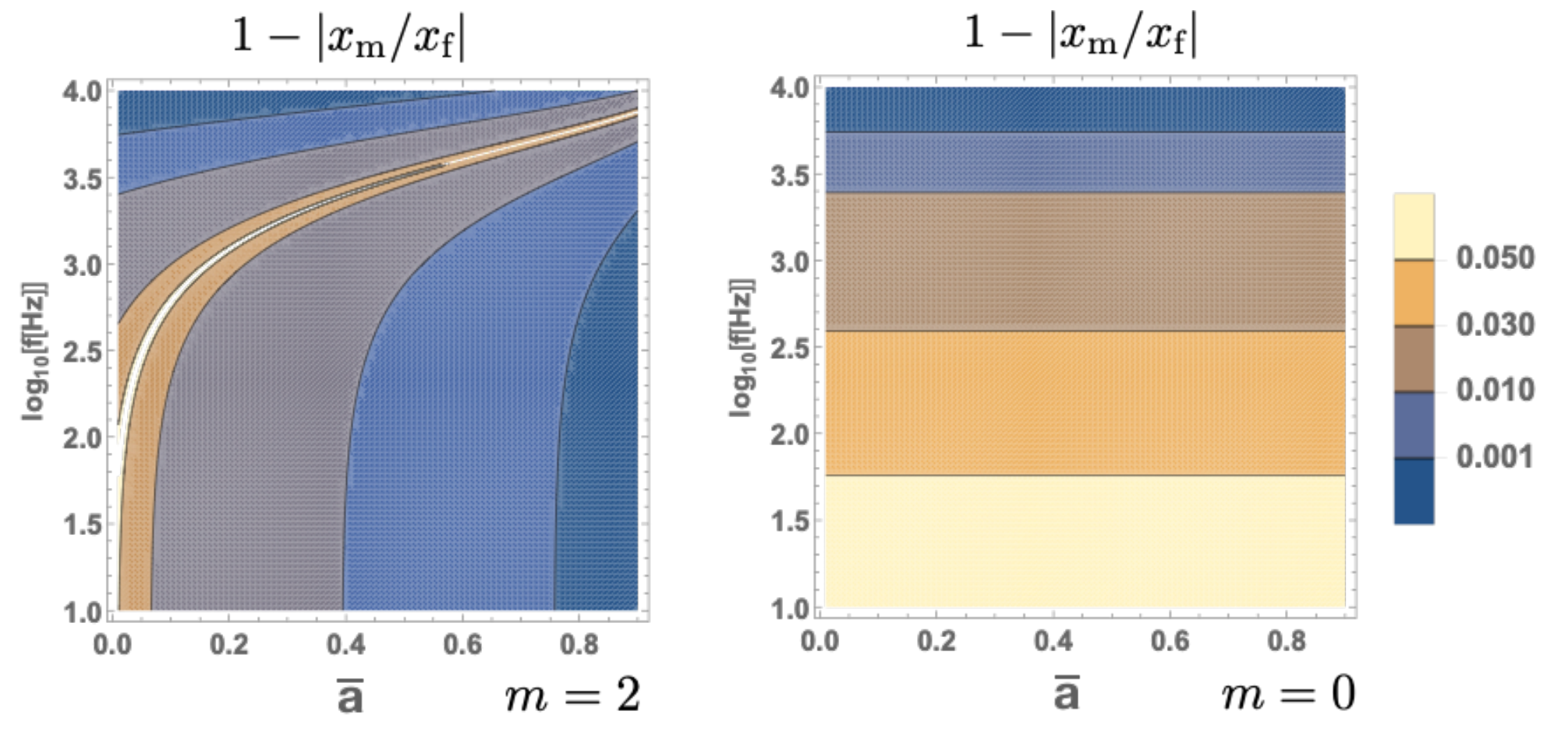}
\caption{Relative difference between the echo reflection points in tortoise coordinates, for ECOs ($x_m$) versus the modified dispersion relation ($x_f$), as a function of frequency for $m=2$ and $0$. The mass of a spinning BH is assumed to be $2.7 M_{\odot}$.
}
\label{fre_mass1}
\end{figure}

To the best of our knowledge, there are two mechanisms to explain the emission of GW echoes. One is the reflection at the surface of ECOs or the mirror angular momentum barrier of a wormhole. Another one is the reflection of highly blue-shifted GWs in the vicinity of the horizon due to a possible Planckian modification in the dispersion relation of GW (FIG. \ref{exo_pla}).
In the former case, the reflection position is at a fixed proper microscopic distance (e.g., a Planck length) outside the would-be horizon. In this case, the approximate point of reflection is given by $x_0 = x_{\text{m}}$ in (\ref{x_m}) in tortoise coordinate [top line in Eq. \ref{x_m}]. On the other hand, in the latter case the reflection position may depend on the frequency of incoming GWs because the reflection due to the Planckian modification in the dispersion relation would take place when the frequency of incoming GWs reaches the Planckian frequency near the horizon.
Therefore, lower frequency incoming GWs can approach closer to the horizon until they reach the Planckian frequencies. In this case, therefore, the position of the reflection surface may be given by $x_0 = x_{\text{f}}$ in (\ref{x_m}). An interesting question is how this difference in the echo mechanisms changes the echo spectrum. Using two different reflection positions given in (\ref{x_m}), the difference between them is shown in FIG. \ref{fre_mass1}, where $1-|x_{\text{m}}/x_{\text{f}}|$ is plotted for $m=0$ and $2$ as a function of the spin $a$ and frequency of incoming GWs $f$.
\begin{figure}[t]
\includegraphics[width=0.5\textwidth]{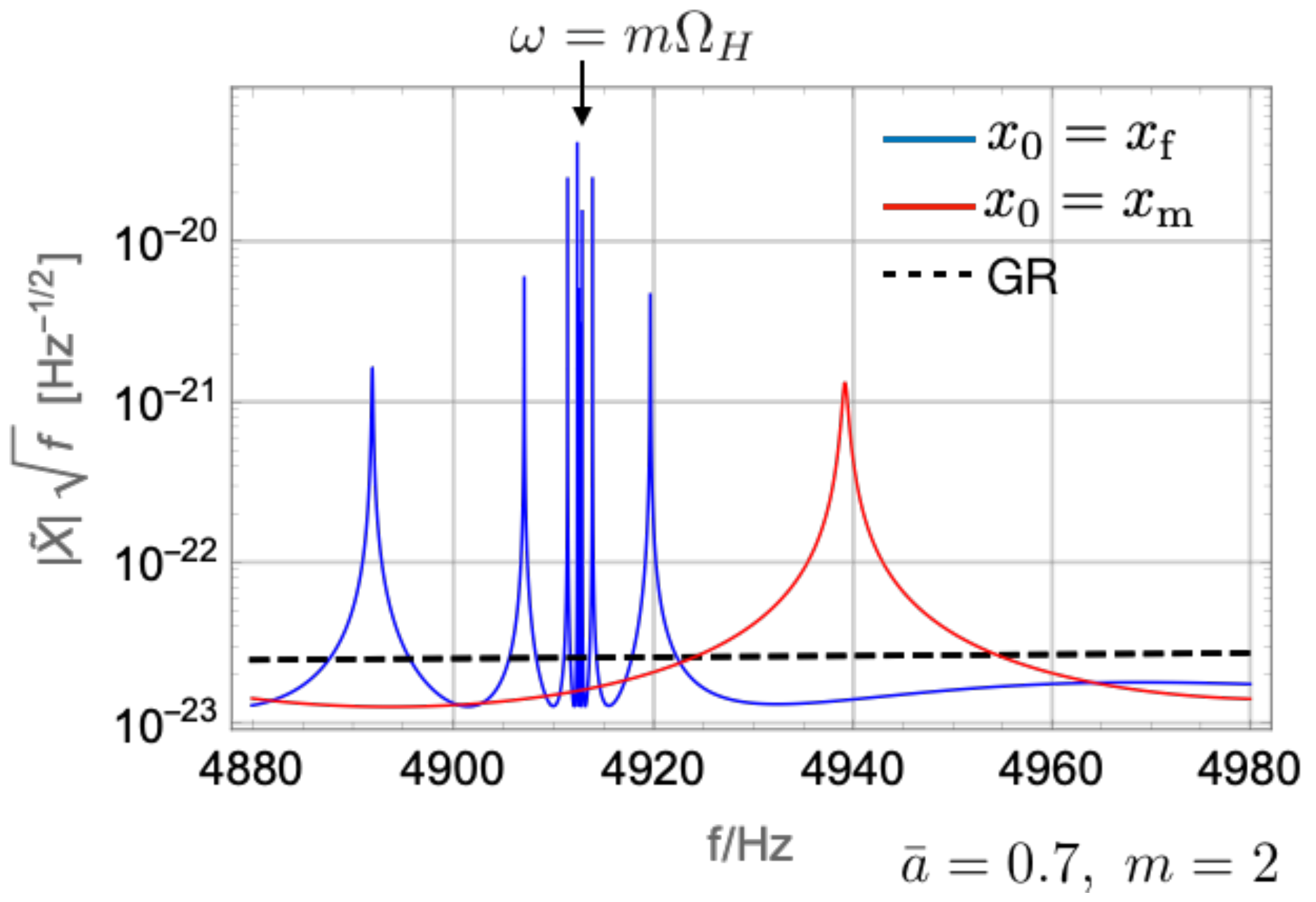}
\caption{Plot of the echo spectra ($\ell=m=2$) near the horizon frequency for $x_0 = x_{\text{m}}$ (red) and $x_{\text{f}}$ (blue). Here we set $M=2.7 M_{\odot}$, $\bar{a} = 0.7$, $\epsilon_{\rm rd} = 0.04$, $D_o = 40$ Mpc, and $\theta = 90^\circ$. We use the Boltzmann reflectivity model with $T_{\rm H} / T_{\rm QH} = 0.6$ and $\gamma = 1$.
}
\label{enlarge_spe}
\end{figure}
\begin{figure}[t]
\includegraphics[width=1\textwidth]{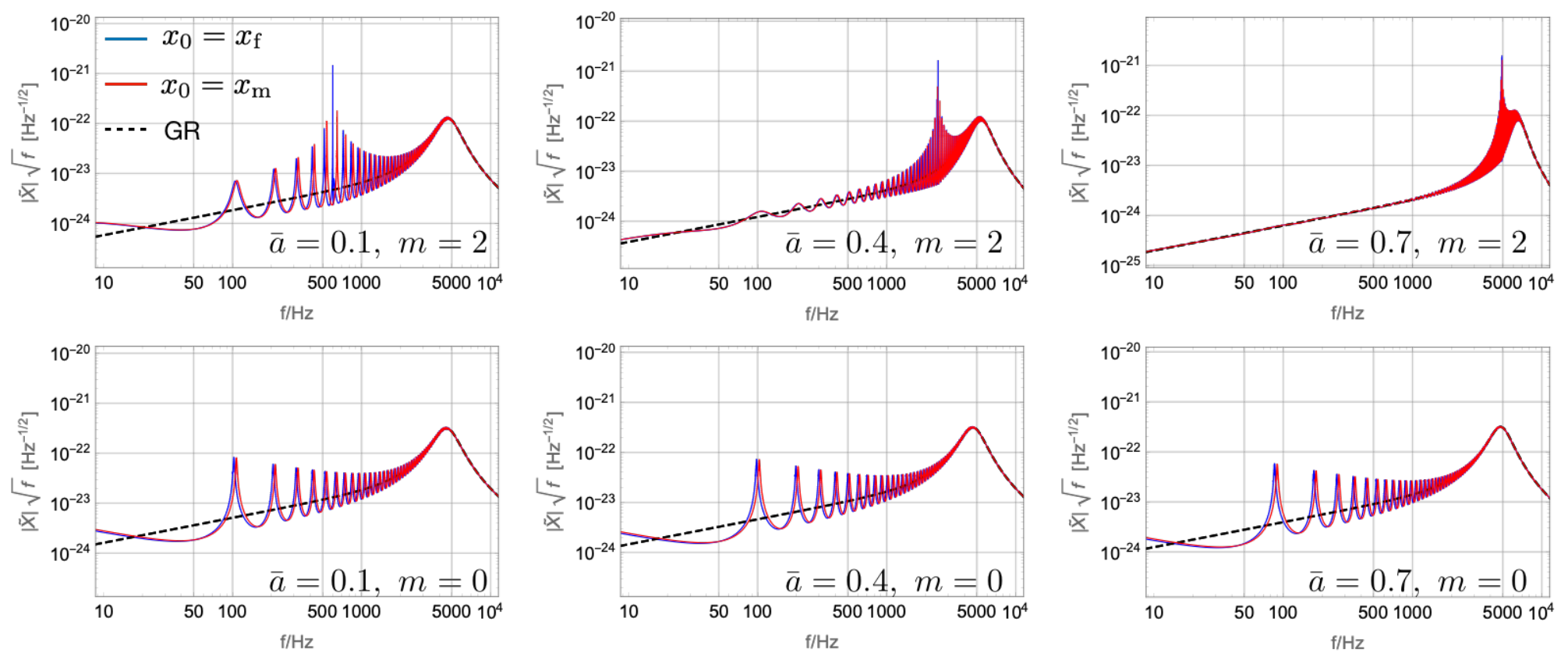}
\caption{Echo spectra for $x_0 = x_{\text{m}}$ (red) and $x_0 = x_{\text{f}}$ (blue) in the Boltzmann reflectivity model with $T_{\rm H} / T_{\rm QH} =0.6$. Here we take $\bar{a} = 0.1$, $0.4$, and $0.7$ with $m=2$ and $m=0$. Other parameters are same as in FIG. \ref{enlarge_spe}.
}
\label{fre_mass2}
\end{figure}
In both cases, one can see that there is small dependence on frequency although $x_0=x_{\text{m}}$ with $m=2$ becomes large in the very vicinity of the horizon frequency $m \Omega_H$. That is, the number density of peaks in frequency space becomes higher at $\omega \simeq m \Omega_H$ (see FIG. \ref{enlarge_spe}), compared to the case of $x_0=x_{\text{m}}$. However, in the tentative detection of echoes following GW170817 \cite{Abedi:2018npz} (which we shall discuss in a forthcoming paper \cite{Oshita:seis2}), the relevant frequency is around $\sim 100$ Hz but the horizon frequency is larger than $1000$ Hz for $\bar{a} \gtrsim 0.2$. Therefore, in the following, we only investigate the case of $x_0 = x_{\text{m}}$. The echo spectra for $x_0 = x_{\text{m}}$ and $x_0 = x_{\text{f}}$, based on the Boltzmann reflectivity model, are shown in FIG. \ref{fre_mass2} for comparison.

\subsection{Spin dependence of echo spectrum}
Here, we study the spin dependence of echo spectra calculated by using the transfer function introduced in the previous section.
\begin{figure}[t]
\includegraphics[width=0.8\textwidth]{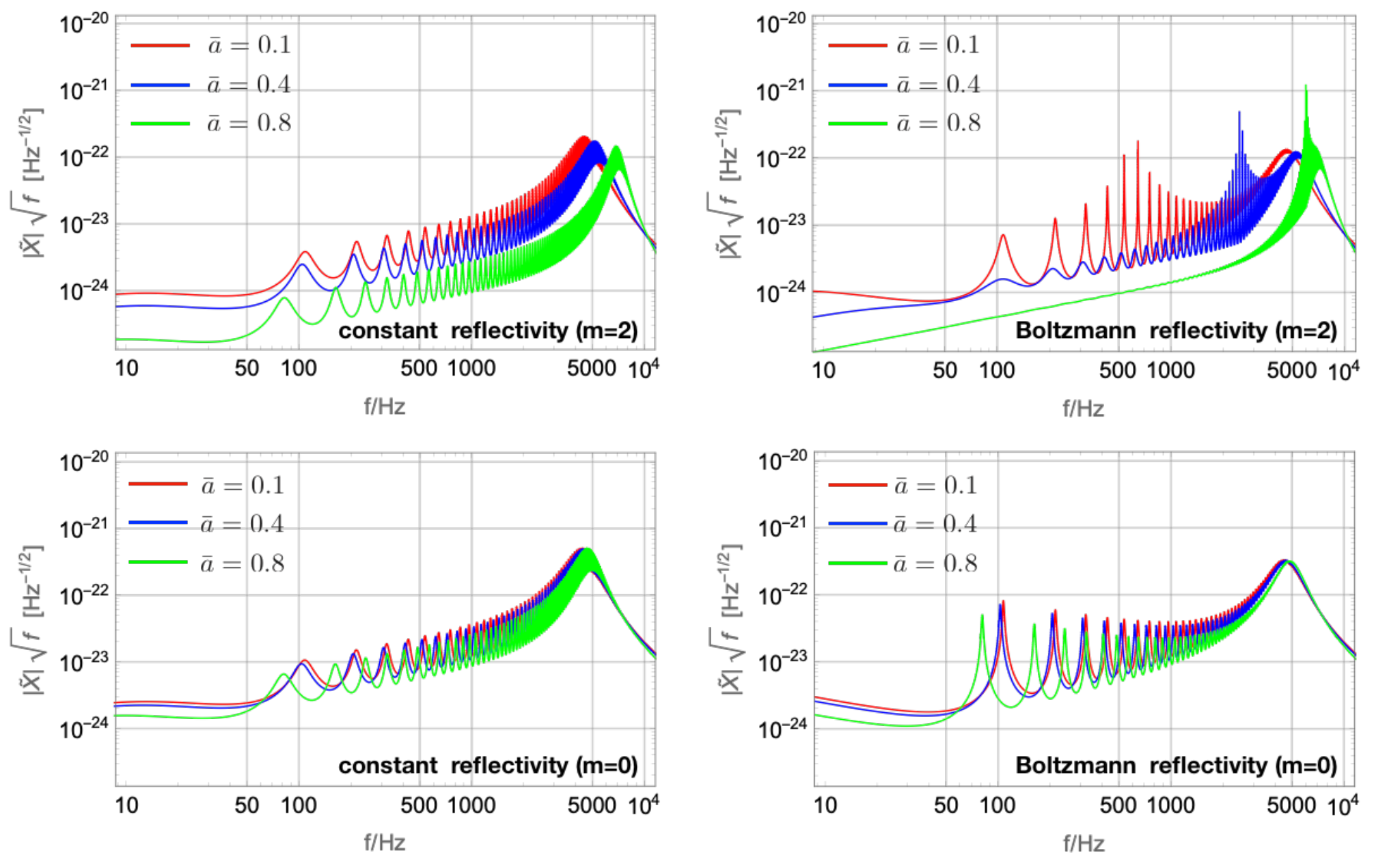}
\caption{Spectra for GW echoes (upper: $\ell = m = 2$, lower: $\ell = 2$, $m=0$) from a remnant BH of $M = 2.7 M_{\odot}$ and $\bar{a}=0.1$ (red), $0.4$ (blue), and $0.8$ (green) in the constant reflectivity model with $R_c = 0.5$ (left) and Boltzmann reflectivity model with $T_{\rm H} / T_{\rm QH} =0.6$ and $\gamma =1$ (right). We set $\theta = 90^\circ$, $D_o = 40$ Mpc, and $\epsilon_{\text{rd}} = 0.04$.}
\label{spin_echo_constant}
\end{figure}
The spin dependence of echo spectra in the constant and Boltzmann reflectivity models with $m=2$ are shown in FIG. \ref{spin_echo_constant} with a fixed energy of ringdown GWs. One can read that a highly spinning BH leads to smaller amplitude of echoes at low frequency. This is because the least-damped QNM frequency is higher for a highly spinning BH and so the amplitude at low frequency is suppressed due to the energy conservation. The amplitude of GW echoes for the Boltzmann reflectivity model is amplified compared to the former case. The echo amplitude at $\omega \sim m \Omega_H$ is highly excited since the reflectivity at the frequency is almost unity (FIG. \ref{spin_echo_constant}).
We also consider the case where an $\ell =2, m=0$ mode dominated the ringdown GWs. Such a situation may occur, for example, after the long-lived remnant of a neutron star merger became almost spheroidal before collapsing into a BH. The spin dependence of echo spectra with $m=0$ is also shown in FIG. \ref{spin_echo_constant}, which has small spin dependence compared to the case of $m=2$. This is because the least damping QNM of $m=0$ has less dependence on the spin of BHs.

\subsection{Importance of overtone in the amplitude of GW echoes}
\label{sec:overtone}
\begin{figure}[t]
\includegraphics[width=0.9\textwidth]{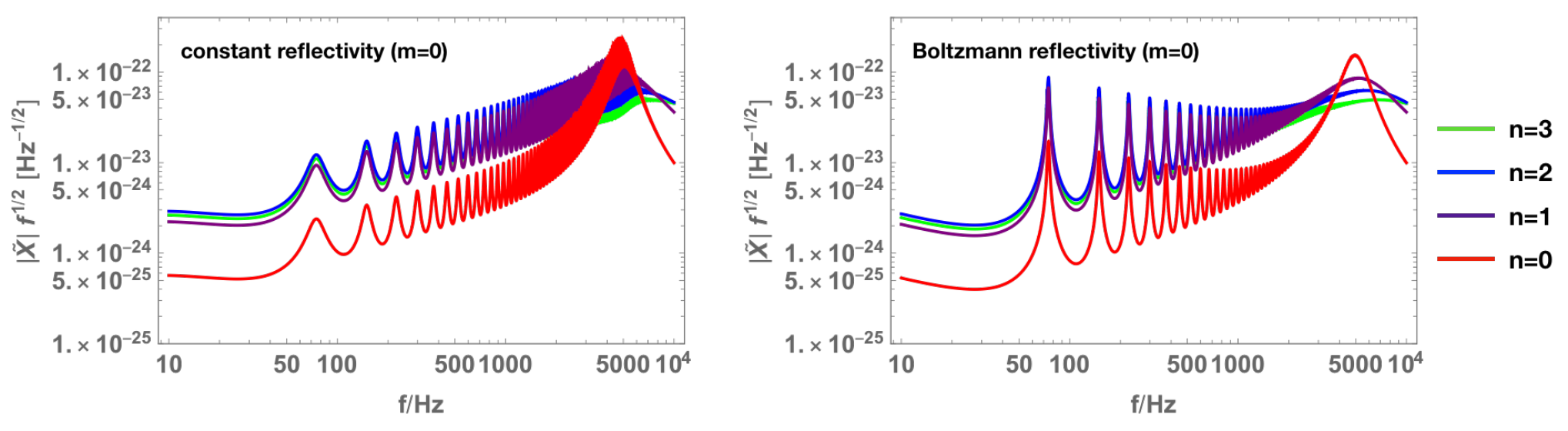}
\caption{Spectra for GW echoes ($\ell = 2$, $m=0$) with a single dominant QNM [$n=0$ (red), $n=1$ (purple), $n=2$ (blue), and $n=3$ (green)] from a remnant BH of $M=2.7 M_{\odot}$ and $\bar{a} = 0.85$ in the constant reflectivity model with $R_c = 0.5$ (left) and the Boltzmann reflectivity model with $T_{\rm H} / T_{\rm QH} = 0.6$ and $\gamma = 1$ (right). We set $\theta = 40^\circ$, $D_o = 40$ Mpc, and $\epsilon_{\rm rd} = 0.04$.}
\label{overtone}
\end{figure}

The reported tentative detection of echoes in GW170817 \cite{Abedi:2018npz} uses the GW observations at $f \lesssim 1000$ Hz, which is well below the (real part of) frequency of the least damping QNM $\sim 5000$ Hz. We here show that if the overtone QNMs are dominant when the ringdown phase starts,\footnote{Recently, it was also pointed out \cite{Giesler:2019uxc} that the overtone QNMs ($n\sim 4$) may dominate the early ringdown in the numerical simulations of binary BHs \cite{Mroue:2013xna}.} the amplitude of GW echoes are highly enhanced at this low-frequency region (FIG. \ref{overtone}).
\footnote{In the Boltzmann reflectivity model with $\ell = m = 2$, the reflectivity at the would-be horizon is highly suppressed in the low-frequency region and so the enhancement of echo amplitude with $m=0$ is more significant than the case of $m=2$.}
Although a detailed comparison with the claimed echoes of GW170817 \cite{Abedi:2018npz} will be deferred to a forthcoming paper \cite{Oshita:seis2}, we explain here why the overtone effect enhances the amplitude of GW echoes in the low-frequency region. Since the overtone QNMs lead to the sharp wave packet in the time domain, it becomes broad in the frequency domain. Therefore, it can highly enhance the amplitude in the low frequency compared to the case where the least damping QNM dominates the early ringdown.

\subsection{Phase shift at the would-be horizon and the invariance of ${\cal R} {\cal R}_{\text{BH}}$ under the generalized Darboux transformation}
\label{sec:phase_shift}
\begin{figure}[t]
    \includegraphics[width=0.6\textwidth]{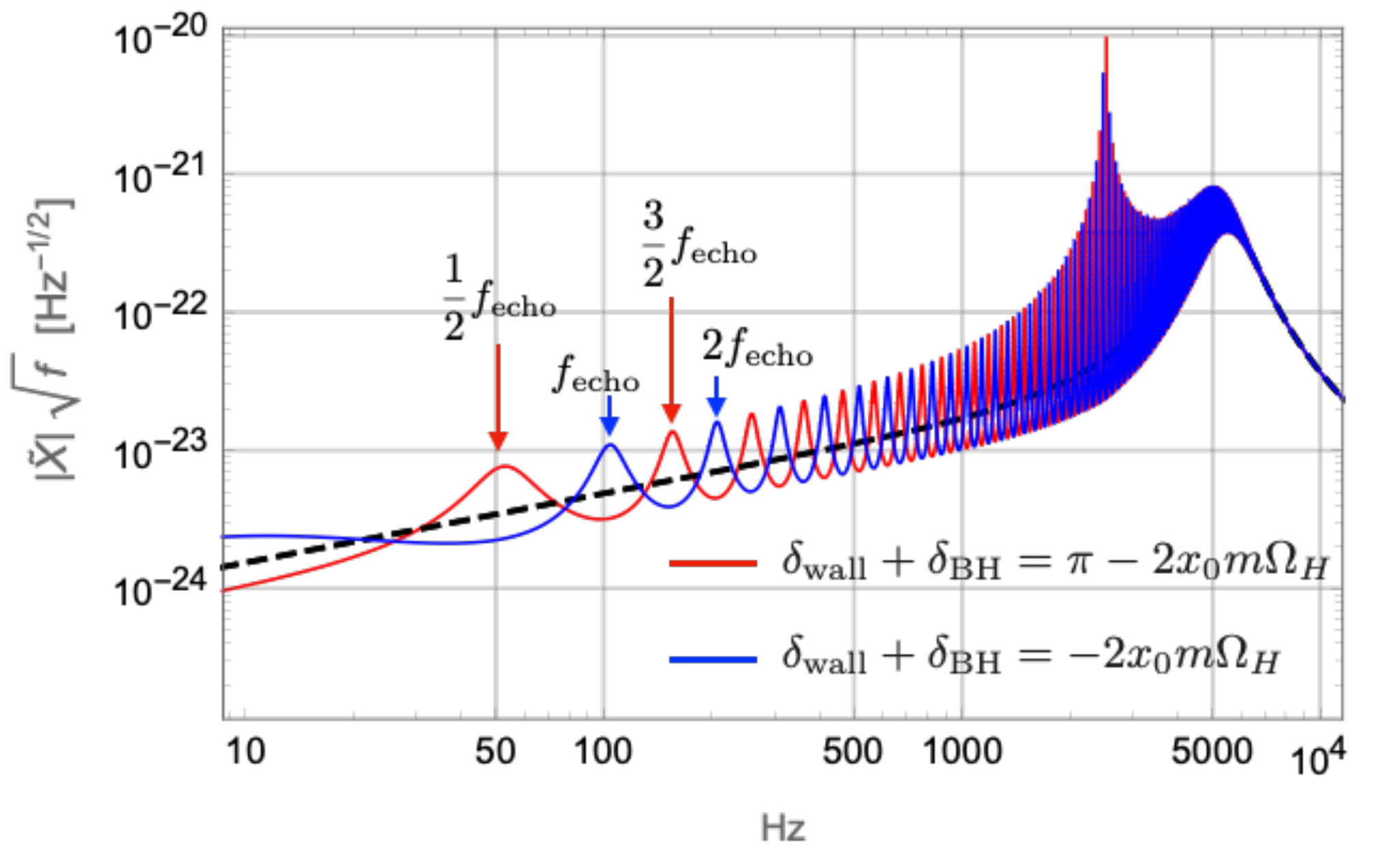}
\caption{Spectra for GW echoes ($\ell = m = 2$) from a remnant BH of $\bar{a}=0.4$ with $\delta_{\text{wall}}+\delta_{\text{BH}} = -2x_0 m \Omega_H$ (blue; consistent with the QNM structure reported for GW170817 in \cite{Abedi:2018npz}) and $\pi-2x_0 m \Omega_H$ (red) in the Boltzmann reflectivity model with $T_{\rm H} / T_{\rm QH} = 1$ and $\gamma =1$. The dashed line shows the ringdown spectrum with the same parameters. We set $\epsilon_{\rm rd} = 0.04$, $M = 2.7 M_{\odot}$, $\theta = 20^\circ$, and $D_o = 40$ Mpc.
}
\label{delta_echo}
\end{figure}

The phase shift at the would-be horizon, $\delta_{\text{wall}} = \text{arg}[{\cal R}]$, is one of the important components to determine the structure of the echo spectrum since the denominator of ${\cal K}$, which has the form of $1-{\cal R} {\cal R}_{\text{BH}} e^{-2i \tilde{\omega} x_0}$, determines the peaks of the echo amplitude in frequency space. One can rewrite the denominator as
\begin{align}
\begin{split}
|1-{\cal R} {\cal R}_{\text{BH}} e^{-2 i \tilde{\omega} x_0}|
&= |1-|{\cal R}{\cal R}_{\text{BH}}| e^{-2 i \tilde{\omega} x_0+i \delta_{\text{wall}}+i \delta_{\text{BH}}}|\\
&= \sqrt{1-2 |{\cal R}{\cal R}_{\text{BH}}| \cos{(2\tilde{\omega} x_0- \delta_{\text{wall}}-\delta_{\text{BH}})} + |{\cal R}{\cal R}_{\text{BH}}|^2}\\
&=\sqrt{(1-|{\cal R}{\cal R}_{\text{BH}}|)^2 + 2 |{\cal R}{\cal R}_{\text{BH}}| (1-\cos{(2\tilde{\omega} x_0- \delta_{\text{wall}}-\delta_{\text{BH}})})}.
\end{split}
\end{align}
The second term vanishes when $2\tilde{\omega} x_0 -(\delta_{\text{wall}} + \delta_{\text{BH}}) =2 \pi n$, where $n$ is an integer.
Therefore, the peaks in the spectrum for GW echoes $f_{\text{peak}}$ exist at
\begin{equation}\displaystyle
f_{\text{peak}} = \frac{\omega}{2 \pi} = 
 \left( n + \frac{\delta_{\text{wall}}+\delta_{\text{BH}}}{2 \pi} \right) f_{\text{echo}}+ \frac{m \Omega_H}{2 \pi},
\end{equation}
where $f_{\text{echo}} \equiv 1/|2 x_0|$, which coincides with the real parts of QNMs [see Eq. \ref{re_omega}].
This means that the phase shift at the would-be horizon can be probed by measuring the echo spectral peaks in frequency space, which does not depend on the reflectivity near the horizon \cite{Conklin:2017lwb}. In FIG. \ref{delta_echo}, we plot the echo spectra for $\delta_{\text{wall}}+\delta_{\text{BH}} = -2 x_0 m \Omega_H$ and $\delta_{\text{wall}}+\delta_{\text{BH}} = \pi-2 x_0 m \Omega_H$.
\begin{figure}[b]
    \includegraphics[width=0.9\textwidth]{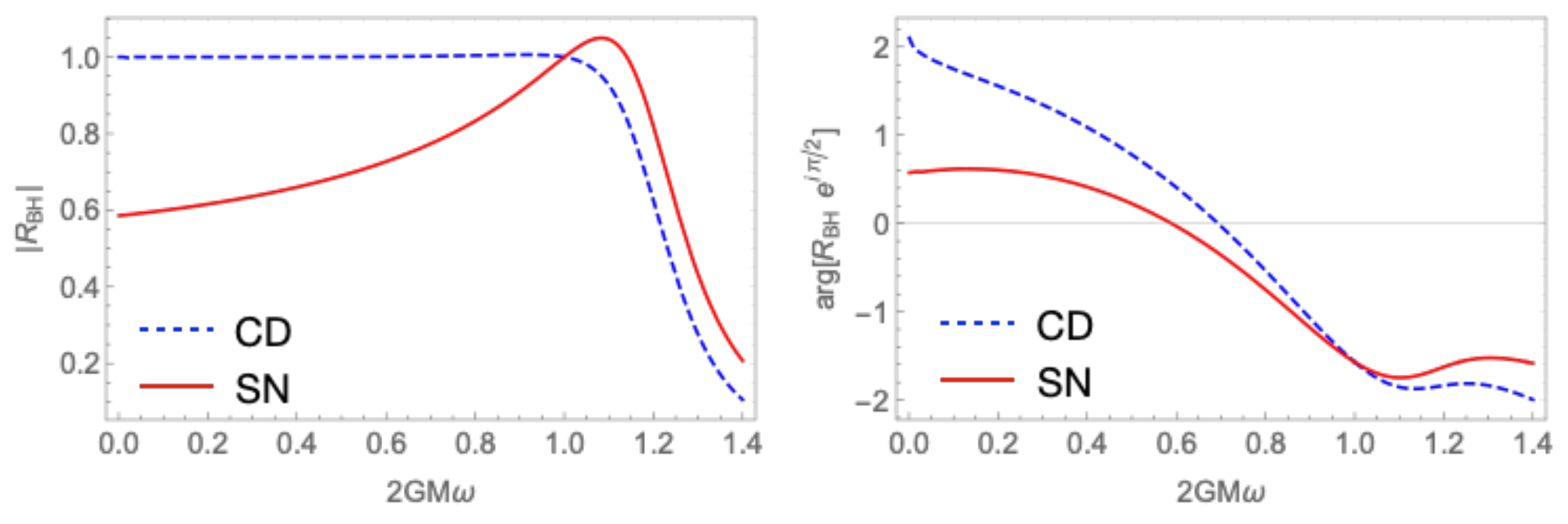}
\caption{The absolute value and argument of the reflection coefficient with $\bar{a} = 0.8$ and $\ell = m = 2$ in the CD (blue dashed) and SN (red solid) equations.
}
\label{ref_cd_sn}
\end{figure}
However, in general, the reflection coefficient ${\cal R}_{\text{BH}}$ depends on the variable of the wave equation. As an example, we plot the reflection coefficient of the mode function for the CD and SN equations in FIG. \ref{ref_cd_sn}, which shows that the reflection coefficients of CD and SN equations, ${\cal R}_{\text{BH}}^{(CD)}$ and ${\cal R}_{\text{BH}}^{(SN)}$, respectively, differ significantly.
However, one can show that ${\cal R} {\cal R}_{\text{BH}}$ is invariant under the generalized Darboux transformation. The generalized Darboux transformation relates a variable $y$ to $Y$ by the following relation
\begin{equation}
Y = \xi (r^{\ast}) y + \zeta (r^{\ast}) \frac{dy}{dr^{\ast}},
\label{trans_GDT}
\end{equation}
and both variables $y$ and $Y$ satisfy the following canonical wave equations:
\begin{equation}
\left( \frac{d^2}{dr^{\ast} {}^2} + \omega^2 - v(r^{\ast}) \right) y =0, \ \ 
\left( \frac{d^2}{dr^{\ast} {}^2} + \omega^2 - u(r^{\ast}) \right) Y =0,
\end{equation}
where $\xi$ and $\zeta$ are transformation functions, and $v$ and $u$ are localized potential barriers. Let us assume that the transformation functions satisfy $\displaystyle \lim_{r^{\ast} \to - \infty} \xi \pm i \tilde{\omega} \zeta = C_{\pm} f_{\pm} (r^{\ast})$, where $C_{\pm}$ is a constant and $f_{\pm} (r^{\ast})$ is a function which does not converge to a finite constant in $r^{\ast} \to -\infty$.\footnote{For example, when the variables $y$ and $Y$ are associated with the variables of the CD and Teukolsky equations, respectively, $f_+ = \Delta^2$ and $f_- = 1$ for a spin-$(-2)$ field.} We also assume the asymptotic form of $y$ as
\begin{equation}\displaystyle
\lim_{r^{\ast} \to -\infty} y = A_{\text{in}} e^{-i \omega r^{\ast}} + A_{\text{out}} e^{i \omega r^{\ast}},
\end{equation}
where $A_{\text{in}}$ and $A_{\text{out}}$ are constants. In this case, the form of $Y$ in the limit of $r^{\ast} \to - \infty$ is given by
\begin{align}\displaystyle
\begin{split}
\lim_{r^{\ast} \to -\infty} Y &= 
A_{\text{in}} (\xi -i \tilde{\omega} \zeta) e^{-i\omega r^{\ast}} + A_{\text{out}} (\xi + i \tilde{\omega} \zeta) e^{i \omega r^{\ast}}\\
&= A_{\text{in}} C_- f_- (r^{\ast}) e^{-i\omega r^{\ast}} + A_{\text{out}} C_+ f_+ (r^{\ast}) e^{i\omega r^{\ast}}.
\end{split}
\end{align}
Therefore we obtain the reflection coefficient with the variable $Y$ as
\begin{equation}
{\cal R}_{\text{BH}}^{(Y)} \equiv \frac{A_{\text{in}} C_-}{A_{\text{out}} C_+} = {\cal R}_{\text{BH}}^{(y)} \frac{C_-}{C_+}.
\label{SNCD1}
\end{equation}
Not only the reflection coefficient, but also the boundary condition should be also transformed by the generalized Darboux transformation. Let us suppose that in the original expression the boundary condition is given by
\begin{equation}\displaystyle
\lim_{r^{\ast} \to - \infty} y = e^{-i\omega r^{\ast}} + {\cal R}^{(y)} e^{i \omega r^{\ast}}.
\label{CD_boundary}
\end{equation}
The transformation (\ref{trans_GDT}) gives the corresponding boundary condition of (\ref{CD_boundary}) in the expression of $Y$ as
\begin{equation}\displaystyle
\lim_{r^{\ast} \to - \infty} Y = (\xi - i \omega \zeta) e^{-i \omega r^{\ast}} + {\cal R}^{(y)} (\xi +i \omega \zeta) e^{i\omega r^{\ast}} = C_- f_- e^{-i\omega r^{\ast}} + {\cal R}^{(y)} C_+ f_+ e^{i\omega r^{\ast}},
\end{equation}
which is equivalent to the boundary condition of
\begin{align}\displaystyle
&\lim_{r^{\ast} \to - \infty} Y = f_- e^{-i \omega r^{\ast}} + {\cal R}^{(Y)} f_+ e^{i\omega r^{\ast}},\\
&\text{with} \ {\cal R}^{(Y)} \equiv {\cal R}^{(y)} \frac{C_+}{C_-}. \label{SNCD2}
\end{align}
From (\ref{SNCD1}) and (\ref{SNCD2}), we finally obtain
\begin{equation}
{\cal R}_{\text{BH}}^{(Y)} {\cal R}^{(Y)} = {\cal R}_{\text{BH}}^{(y)} {\cal R}^{(y)},
\end{equation}
which means that the QNMs of GW echoes are invariant under the generalized Darboux transformation.

\begin{figure}[t]
    \includegraphics[width=0.6\textwidth]{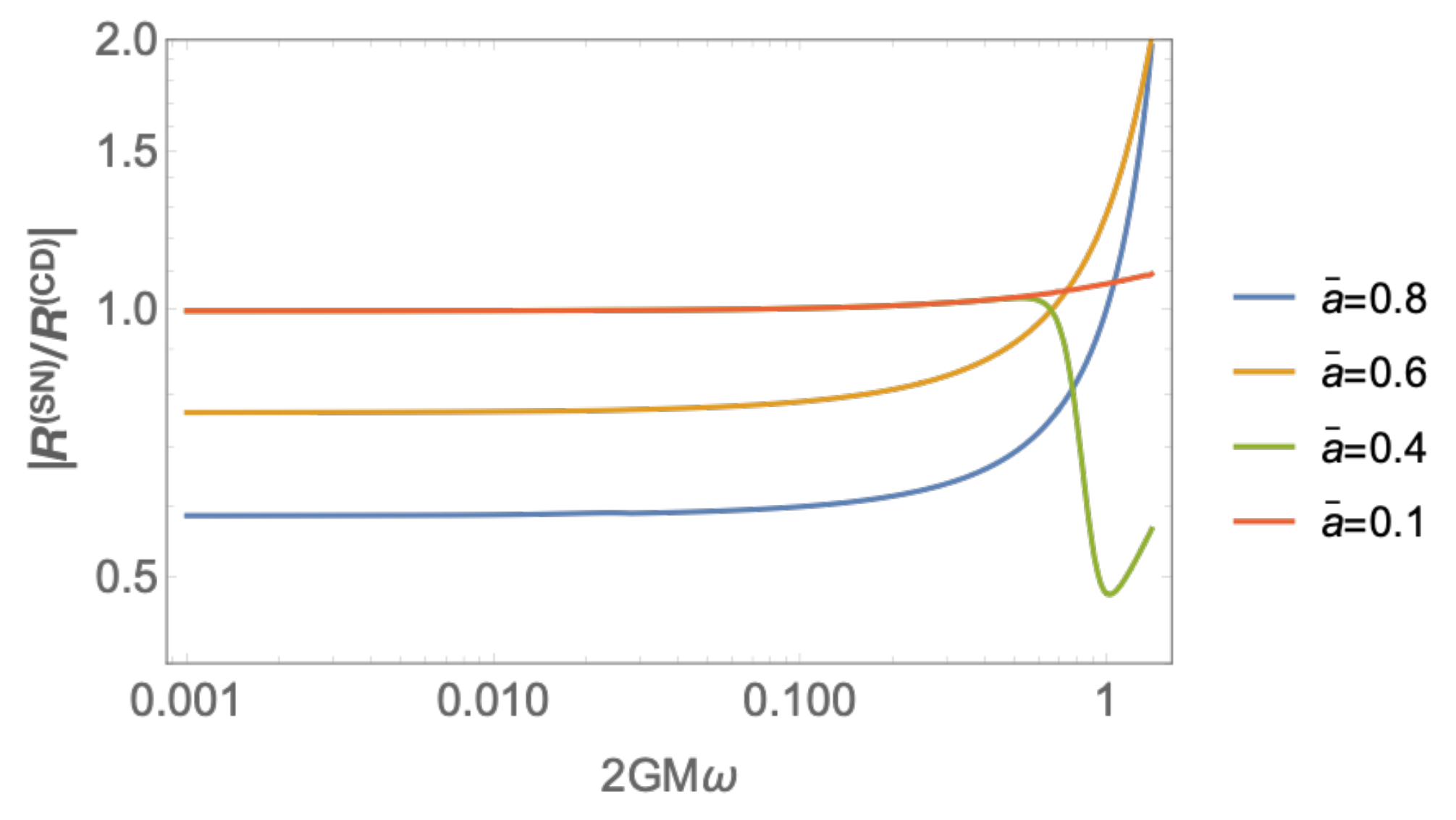}
\caption{Plots of ${\cal R}^{(SN)}/{\cal R}^{(CD)}$ for $\bar{a} = 0.1$, $0.4$, $0.6$, and $0.8$ with $\ell = m= 2$.
}
\label{CD_SN_comp}
\end{figure}

On the other hand, this also means that the value of reflectivity imposed at $r^{\ast}= x_0$ depends on a variable of linear perturbations. To give an example, we show the frequency dependence of the ratio ${\cal R}^{(y)}/{\cal R}^{(Y)} = {\cal R}_{\text{BH}}^{(Y)} / {\cal R}_{\text{BH}}^{(y)}$ for CD and SN variables in FIG. \ref{CD_SN_comp}. One can find out that the value of reflectivity depends on which variable is chosen and so when we use the SN or Teukolsky equations, whose reflection coefficients calculated from mode functions are not equivalent to the root of energy reflectivity, one has to properly transform the boundary condition.

\section{Conclusion}
We have provided a detailed framework for calculating GW echo emission from spinning BHs based on the CD and SN equations, which can be applied for physical reflectivities expected from quantum event horizons.
Using this framework:
\begin{enumerate}
    \item 
 We have put constraints on the reflectivity of quantum BHs, to avoid ergoregion instability up to the Thorne limit $\bar{a} \leq 0.998$, yielding a constraint on the energy reflectivity, $R_c \lesssim 0.72$, for the constant reflectivity model. For the (generalized) Boltzmann reflectivity model, we find the upper bound for the quantum horizon temperature of $T_{\rm QH} \lesssim 1.86 \times T_{\rm H}$.
\item We also investigated how the spectrum depends on the assumed constant/Boltzmann reflectivity model, the spin of the BH, and the phase shift at the would-be horizon. 
The echo amplitude for the Boltzmann reflectivity model is larger than that for the constant reflectivity model for any spin and for both $m=2$ and $m=0$ modes.
\item We pointed out that the amplitude of GW echoes in the low-frequency regime can be highly enhanced when the overtone QNMs dominate the early ringdown phase. For a BH mass of $\sim 3M_\odot$ this regime is around $100$ Hz, which is relevant for ground-based detectors. A detailed study of how overtones can affect the observability of echoes following BH formation will be presented in a forthcoming paper \cite{Oshita:seis2}.
\item We investigated GW spectra with two different echo mechanisms, i.e. the ECO scenario and the modified dispersion relation scenario. The former scenario gives the mass-dependent reflection radius $r^{\ast}=x_{\text{m}}$ and the latter one gives the frequency-dependent reflection radius $r^{\ast}=x_{\text{f}}$. We found out that the difference in echo spectra can be seen in the very vicinity of the horizon frequency, where the number density of echo peaks in the frequency space is higher for $r^{\ast} =x_{\text{f}}$. In other frequency regions, both spectra match very well.
\item The locations of echo peaks in frequency space depend on the phase shift at the would-be horizon $\delta_{\text{wall}}$. For example, $\delta_{\text{wall}} + \delta_{\text{BH}} = -2x_0 m \Omega_H$ and $\pi-2x_0 m \Omega_H$ give the echo peaks located at $n f_{\text{echo}}$ and at $(n+1/2) f_{\text{echo}}$, respectively. The tentative detection of echoes in GW170817, if real, would be consistent with the former case \cite{Abedi:2018npz}.

\item Finally, we also found the invariance of ${\cal R} {\cal R}_{\text{BH}}$ under the generalized Darboux transformation, which means that the echo QNMs are invariant under the transformation and are genuine covariant observables.

\end{enumerate}

The differences in the spectral features between different mechanisms can be probed by future observations. By applying the methodology outlined in this work, we will discuss in a forthcoming paper \cite{Oshita:seis2} GW echo signals from astrophysical stellar collapses, such as binary neutron stars and failed supernovae.

\begin{acknowledgements}
We thank all the participants in our weekly group meetings for their patience during our discussions. D. T. thanks the Perimeter Institute for their hospitality during his visit. This work was supported by the University of Waterloo, Natural Sciences and Engineering Research Council of Canada (NSERC), and the Perimeter Institute for Theoretical Physics. N. O. is supported by the JSPS Overseas Research Fellowships. D. T. is supported by the Advanced Leading Graduate Course for Photon Science (ALPS) at the University of Tokyo, and by JSPS KAKENHI Grant No JP19J21578. Research at the Perimeter Institute is supported by the Government of Canada through Industry Canada, and by the Province of Ontario through the Ministry of Research and Innovation.
\end{acknowledgements}

\appendix

\section{Explicit for of the functions $\rho$, $\Xi_{i}$, $\Theta_{ij}$, and $\kappa$}
\label{app:ex}
In this appendix, we provide the exact expressions for some functions in the CD equation.
\begin{align}
\rho^2 &\equiv r^2+a^2-am/\omega,\\
\Xi_{i} & \equiv \frac{\Delta^2}{\rho^8} (F+b_i),\\
\Theta_{ij} &\equiv i \omega +\frac{1}{F-b_i} \left( \frac{\Delta}{\rho^2} \frac{dF}{dr} -\kappa_{ij} \right),\\
\begin{split}
\kappa &\equiv (\lambda^2 (\lambda+2)^2 +144 a^2\omega^2 (m-a\omega)^2 -a^2 \omega^2 (40 \lambda^2-48 \lambda)+a\omega m (40 \lambda^2 +48 \lambda))^{1/2}\\
&~~~~+12 i \omega GM,
\end{split}\\
F & \equiv \frac{\lambda \rho^4 +3 \rho^2 (r^2-a^2) -3r^2 \Delta}{\Delta},\\
g & \equiv \lambda \rho^4 +3 \rho^2 (r^2-a^2) -3r^2 \Delta,\\
h & \equiv g' \Delta - g \Delta'.
\end{align}

\section{Calculation of the amplification factor from the SN equation}
\label{app:SN}
\begin{figure}[t]
\includegraphics[width=0.5\textwidth]{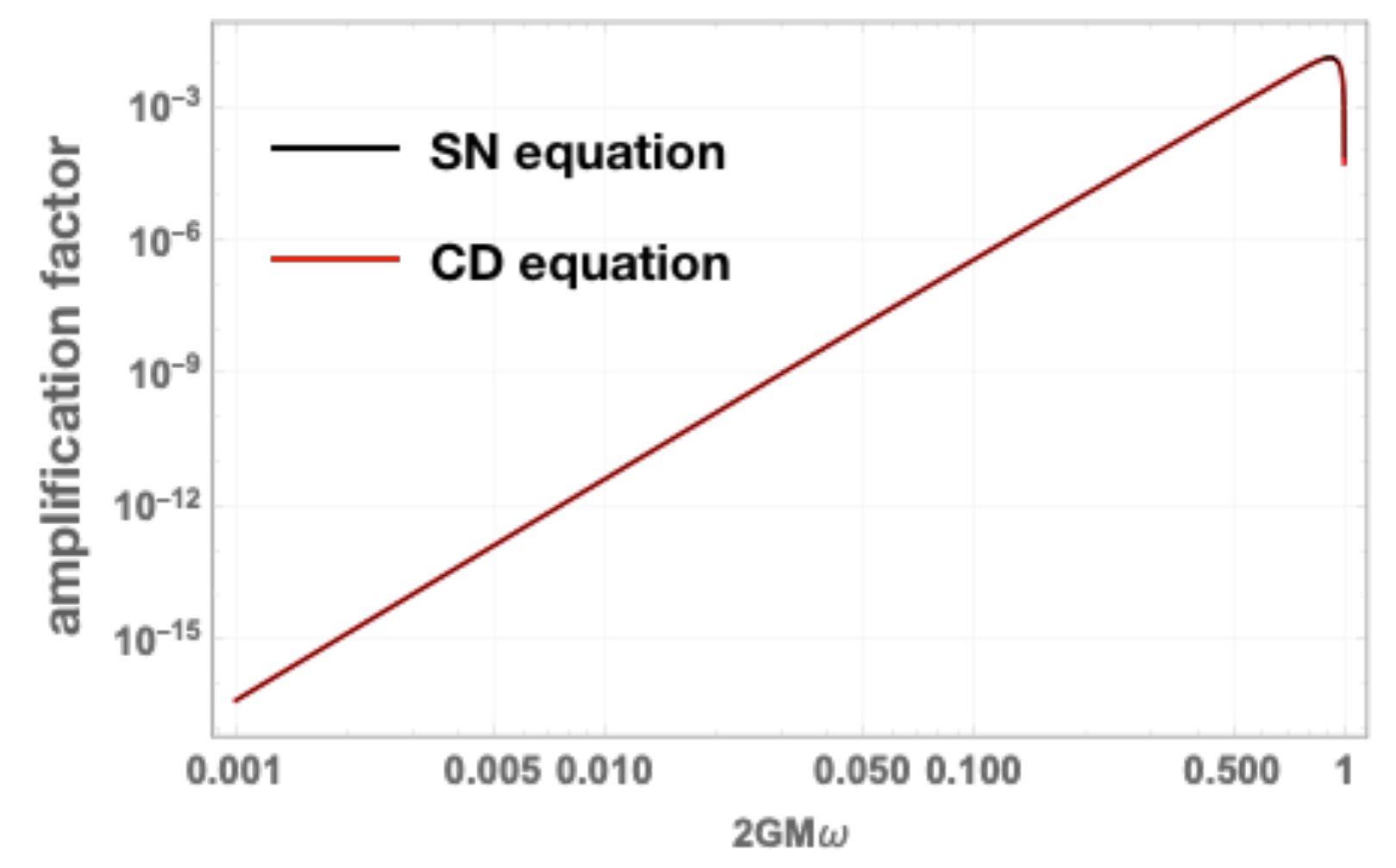}
\caption{The amplification factors calculated in the CD and SN equations for $\bar{a} =0.8$ and $l=m=2$.
}
\label{ampl_cd_sn}
\end{figure}

The amplification factor is an observable quantity and so it does not depend on the variable of perturbation. In this appendix we review the procedures to obtain the amplification factor from the SN equation and compare the amplification factors obtained from the CD and SN equation for a consistency check.
The solution of the homogeneous SN equation has the asymptotic form of
\begin{equation}
{}_s X_{lm} =
\begin{cases}
\tilde{A}_{\text{in}} e^{-i \tilde{\omega} r^{\ast}} + \tilde{A}_{\text{out}} e^{i \tilde{\omega} r^{\ast}} & \text{for} \ r^{\ast} \to - \infty,\\
\tilde{B}_{\text{in}} e^{-i \omega r^{\ast}} + \tilde{B}_{\text{out}} e^{i \omega r^{\ast}} & \text{for} \ r^{\ast} \to  +\infty,
\end{cases}
\end{equation}
and the energy flux spectra near the horizon in terms of the amplitudes $\tilde{A}_{\text{in}}$ and $\tilde{A}_{\text{out}}$ are given by \cite{Conklin:2019fcs}
\begin{equation}
\frac{dE_{\text{out}}}{d\omega} = \frac{8 \omega \tilde{\omega}}{|b_0^2|} |\tilde{A}_{\text{out}}|^2, \ 
\frac{dE_{\text{in}}}{d\omega} = \frac{8 \omega \tilde{\omega}}{|C^2|} |\tilde{A}_{\text{in}}|^2.
\end{equation}
Therefore, one can obtain the energy reflection and transmissivity of the angular momentum barrier
\begin{align}
|{\mathcal I}^{\text{ref}}|^2 &\equiv \frac{|C|^2}{|b_0|^2} \left| \frac{\tilde{A}_{\text{out}}}{\tilde{A}_{\text{in}}} \right|^2,
\label{ref}
\end{align}
where $|C|$ and $b_0$ are given by
\begin{align}
\begin{split}
|C|^2 &\equiv \lambda^4 + 4 \lambda^3 + \lambda^2 (-40 a^2 \omega^2+ 40 a m \omega+4) +48 a\lambda \omega (a \omega+m)\\
&+144 \omega^2 (a^4 \omega^2 -2a^3 m \omega+ a^2 m^2 + 1/4),
\end{split}\\
\begin{split}
b_0 &\equiv \lambda^2 +2 \lambda -96 \tilde{\omega}^2 G^2M^2 +72 \tilde{\omega} GM r_+ \omega-12 \omega^2 r_+^2\\
&-i \left[ 16 \tilde{\omega} GM (\lambda+3-3GM/r_+)-12GM\omega -8 \lambda r_+ \omega \right].
\end{split}
\end{align}
Then we numerically confirmed that both the CD and SN equations give the consistent amplification factor as is shown in FIG. \ref{ampl_cd_sn}.

\end{document}